\documentclass[aps,pre,superscriptaddress,onecolumn,10pt]{revtex4-2}

\usepackage{url}
\usepackage{xcolor}
\usepackage[normalem]{ulem}
\usepackage{graphicx}
\usepackage{subfigure}
\usepackage{amssymb}
\usepackage{booktabs}
\usepackage{array} 
\usepackage{tabularx}
\usepackage{bm}
\usepackage{amsmath}
\usepackage{bbm}
\usepackage[utf8]{inputenc}
\usepackage{comment}
\usepackage{braket}

\usepackage{hyperref}
\hypersetup{colorlinks=true,linkcolor=blue,citecolor=blue,urlcolor=blue}

\begin{document}

\title{Energetics of Fano coherence generation}

\author{Ludovica Donati}
\affiliation{Istituto Nazionale di Ottica del Consiglio Nazionale delle Ricerche (CNR-INO), Largo Enrico Fermi 6, I-50125 Firenze, Italy.}
\affiliation{European Laboratory for Non-linear Spectroscopy, Università di Firenze, I-50019 Sesto Fiorentino, Italy.}
\affiliation{Dipartimento di Fisica e Astronomia, Università di Firenze, via Sansone 1, I-50019 Sesto Fiorentino, Italy.}

\author{Francesco Saverio Cataliotti}
\affiliation{Istituto Nazionale di Ottica del Consiglio Nazionale delle Ricerche (CNR-INO), Largo Enrico Fermi 6, I-50125 Firenze, Italy.}
\affiliation{European Laboratory for Non-linear Spectroscopy, Università di Firenze, I-50019 Sesto Fiorentino, Italy.}
\affiliation{Dipartimento di Fisica e Astronomia, Università di Firenze, via Sansone 1, I-50019 Sesto Fiorentino, Italy.}

\author{Stefano Gherardini}
\affiliation{Istituto Nazionale di Ottica del Consiglio Nazionale delle Ricerche (CNR-INO), Largo Enrico Fermi 6, I-50125 Firenze, Italy.}
\affiliation{European Laboratory for Non-linear Spectroscopy, Università di Firenze, I-50019 Sesto Fiorentino, Italy.}

\begin{abstract}
In a multi-level quantum system Fano coherences stand for the formation of quantum coherences due to the interaction with the continuum of modes characterizing an incoherent process. When the incoherent source vanishes, Fano coherences tend to disappear.  
In this paper we propose a V-type three-level quantum system on which we certify the presence of genuinely quantum traits underlying the generation of Fano coherences.
We do this by determining work conditions that allows for the loss of positivity of the Kirkwood-Dirac quasiprobability distribution of the stochastic energy variations within the discrete system. 
We also show the existence of nonequilibrium regimes where the generation of Fano coherences leads to a non-negligible amount of extractable work, however provided the initial state of the discrete system is in a superposition of the energy eigenbasis. We conclude the paper by studying the thermodynamic efficiency of the whole process.
\end{abstract}

\maketitle

\section{Introduction}

Three-level systems are a cornerstone model to study a plethora of different quantum phenomena, especially in the field of quantum optics and atomic physics~\cite{Scully:Zub,Loudon}. Depending on the system's layout, this model takes into account the presence of coherent superposition of either the lower or upper states in ${\rm \Lambda}$-type or V-type configuration, respectively. For instance, the ${\rm \Lambda}$-type configuration has been used in the analysis of the interaction between two coherent radiation modes and two optical atomic transitions, during which the occurrence of quantum interference in absorption leads to population trapping in a superposition state that does not interact with the laser fields~\cite{Arimondo:Orriols,Gray,Ar:E}. The phenomenon, identified as Coherent-Population Trapping (CPT), causes alterations in the optical properties of the medium. Specifically, the absorption profile for one field undergoes modification due to the presence of the other field, making the medium electromagnetically-induced transparent~\cite{Harris:Field,Harris}. In this scenario the state of the quantum system changes coherently being driven by a coherent (laser) field~\cite{Kozlov:Scully}.

Even more remarkable is the alternative mechanism of inducing quantum coherence by means of \textit{incoherent} sources such as a broadband laser or a thermal source, as well as interactions with the surrounding environment~\cite{Agarwal,HegerfeldtPRA1993}. These sources are distinguished by a continuum of modes, as opposed to a single coherent mode of a laser source. Since the 1990s, there has been a growing interest in generating quantum coherence through incoherent processes, such as spontaneous emission and incoherent pumping. 
Several studies concerning the way a system can interact with vacuum modes, arising during the process of spontaneous emission, have proliferated as detailed later. The primary focus of these investigations was to elucidate the interference phenomena that stem from the existence of multiple and closely spaced emission pathways, leading to a shared ground level within a three-level V-type system. The interference can manifest as quantum beats in the emission radiation intensity~\cite{Scully:Zub,Haroche}, as population inversion~\cite{Gong}, and as the emergence of dark lines and narrower linewidths in the spectrum of spontaneous emission~\cite{Zhu:Narducci,Zhu:Chan,Zhou:Swain} or even its quenching~\cite{Cardimona}. 

Gaining control over the properties of the fluorescence spectrum is difficult when relying solely on the interference of two decay channels. However, the introduction of an additional incoherent mechanism can enhance flexibility in selecting control parameters with a higher degree of freedom, without degrading the coherence of the quantum system as one might expect~\cite{Kapale:Scully,Kozlov:Scully}. 
An example of this incoherent mechanism can be given by an external pumping from a broadband radiation source that simultaneously drives two almost-degenerate states in a three-level system.
Interestingly, the combination of incoherent pumping and spontaneous emission results in the formation of Fano coherences or interferences, which can be stationary or quasi-stationary. These coherences emerge from the interaction between discrete atomic energy levels and the continuum of modes associated with the two incoherent processes~\cite{Koyu:Tscherbul,Dodin:Tscherbul,Koyu:Dodin}. The resilience of coherences under the aforementioned ``noisy'' conditions needed for their generation, which can be achieved also in ${\rm \Lambda}$-type systems~\cite{Ou:Liang,Koyu}, has particular significance for systems in contact with thermal reservoirs, such as quantum heat engines~\cite{ScullyPNAS2011}, or with thermal radiation, as customary in photo-conversion devices~\cite{Svidzinsky}. In particular, in the latter case, Svidzinsky \emph{et al.}~\cite{Svidzinsky} theoretically demonstrate that Fano interferences might enable the mitigation of spontaneous emission, thereby reducing radiative recombination phenomena. 
To show this, the photo-conversion devices (photocell) is modeled with a V-type three-level system driven by incoherent light source, wherein the excited states represent conduction band states decaying into a common valence band state. Thus, quantum coherence between the excited states of the model would theoretically lead to an increase in extractable current from the device. Consequently, this enhancement would boost the output power and conversion efficiency.

Despite extensive research conducted on the topic and evident technological applications, an experiment proving the existence of Fano coherences produced by the interplay of incoherent pumping and spontaneous emission is still missing as far we know. Currently, the atomic platform stands as the most suitable candidate for such measurements, given its capability to finely adjust the parameters that define a three-level V-type system. In \cite{Dodin:Tscherbul} and subsequently in \cite{Koyu:Dodin}, a proposal was outlined for an experiment on a system comprising beams of Calcium atoms excited by a broadband polarized laser within a uniform magnetic field. Moreover, in a magneto-optical trap of Rubidium atoms, enhanced beat amplitudes due to the collective emission of light, akin Fano coherences due to the interaction with the vacuum modes, have been recently detected in \cite{Han:Lee}.

Our paper explores the influence of quantum coherence in a V-type three-level system with optical transitions subjected to a incoherent light source. Specifically, we are going to investigate a system featuring almost-degenerate upper levels, consistent with prior studies~\cite{Koyu:Tscherbul,Dodin:Tscherbul,Koyu:Dodin}, yet deviating from~\cite{Kozlov:Scully,Svidzinsky}. This framework indeed aims to replicate a more realistic system, akin to those achievable in atomic platforms. However, we maintain the same formalism for modelling the dynamics, namely a quantum Markovian master equation in the Schr\"odinger picture. Particular interest will be devoted to the main thermodynamic aspects behind the generation of Fano coherences in the V-type three-level system. 
In this context, the aims of our investigation are the following:\\
(i) To certify that the generation of Fano coherences has genuinely quantum traits. We attain it by observing the loss of positivity (\emph{i.e.}, negative real parts or even non-null imaginary parts) of the Kirkwood-Dirac quasiprobability distribution of the system energies [Sec.~\ref{sec_quasiprobs}]. The latter are evaluated, respectively, at the initial and final times of the transformation under scrutiny that gives rise to Fano coherences [see Sec.~\ref{sec_modelling}].\\
(ii) To optimize both the initial quantum state of the three-level system (before the latter interacts with the light source) and the parameters of the system, including the coupling strength with the light field, such that the non-positivity of some quasiprobabilities is enhanced. This aspect is doubly important: from the one hand, we can determine under which conditions Fano coherence are generated from a process with pronounced genuinely quantum traits; from the other hand, it can lead to a thermodynamic advantage, being the presence of negativity (\emph{i.e.}, some quasiprobabilties have negative real parts) a necessary condition for enhanced work extraction~\cite{hernandez2022experimental}. We perform the task (ii) in Sec.~\ref{sec_thermodynamics}, where we will study to what extent the process generating Fano coherences (\emph{i.e.}, the three-level system illuminated by incoherent light source) can be employed for energy-conversion purposes. Specifically, we are going to compute the amount of {\it extractable work} in the working conditions outlined in Secs.~\ref{sec_modelling} and \ref{sec_certification}, by looking for the parameters' values that maximizes work extraction. In doing this, we will identify the range of parameters values that allow for work extraction due to negativity. We conclude the paper by discussing the thermodynamic efficiency of the whole process.  

\section{Results}

\subsection{Model}\label{sec_modelling}

\begin{figure}[b!]
\centering
\includegraphics[width=0.45\textwidth]{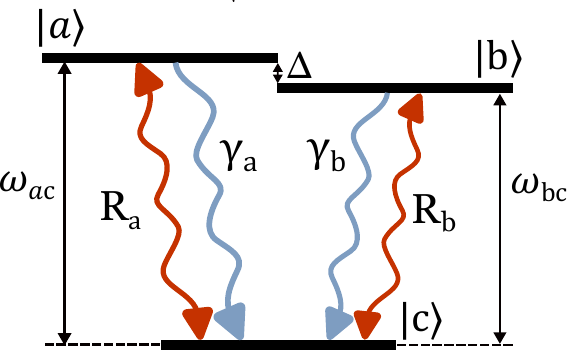}
\caption{The energy level configuration for the V-type three-level system under consideration consists of two nearly degenerate excited levels, denoted as $|a\rangle$ and $|b\rangle$ with a frequency splitting of $\rm \Delta$. These levels are incoherently pumped at rates $R_a$ and $R_b$ respectively, originating from the ground level $|c\rangle$. From the ground level, $|a\rangle$ can decay at rate $\gamma_a$, while $|b\rangle$ can decay at rate $\gamma_b$.}
\label{fig:3LS}
\end{figure}

The V-type three-level system under investigation corresponds to the general configuration depicted in Fig.~\ref{fig:3LS}. In the figure, $|a\rangle$ and $|b\rangle$ are the excited state levels from which the system (\emph{e.g.} an atom) decay on the ground state $|c\rangle$ with rate $\gamma_a,\gamma_b$ respectively. Additionally, both excited states are coupled to the ground via incoherent pumping, (\emph{e.g.}~thermal radiation), with rate $R_a,R_b$. The angular frequencies of the two transitions are indicated as $\omega_{ac} \equiv \omega_a-\omega_c$ and $\omega_{bc} \equiv \omega_b-\omega_c$, while the upper levels splitting as $\rm{\Delta}\equiv\omega_{ac}-\omega_{bc}$. This scenario is thus described by the interaction between the system and the radiation field modelled as a \textit{thermal reservoir}. 

We describe the state of the system via a density operator formalism, as quantum coherences shall be generated during its dynamics. In our setting, as shown below, the quantum dynamics of the system's density operator $\rho_S(t)$ are derived using a quantum Markovian master equation, derived from the Liouville-von Neumann equation governing the density operator $\hat{\rho}(t) = \hat{\rho}_S(t)\otimes\hat{\rho}_R(t)$ of the whole compound comprising the system (S) and the reservoir (R). This model, as well as its solution, has been already discussed in several references so far~\cite{Kozlov:Scully,Koyu:Tscherbul,Koyu:Dodin,Svidzinsky,Dodin:Brumer,Dodin:Brumer2,Petruccione}. However, being sometimes the derivation of the model lacking or not fully explained, we report below a comprehensive derivation that has also pedagogical function.

We start by settling the differential equation for $\hat{\rho}_S(t)$ in the interaction representation, where the Hamiltonian of the whole system is solely determined by the interaction term. For this purpose, we set in the Rotating Wave Aproximation (RWA), under which the interaction Hamiltonian has the following form: 
\begin{equation}\label{eq:interaction_H}
    \hat{H}_I(t) = \hbar\sum_{\lambda=1}^2\sum_{\bf k} g_{\bf{k},\lambda}^a e^{i(\omega_{ac}-\nu_k)t}|a\rangle\langle c|\,\hat{a}_{\bf{k},\lambda}+\hbar\sum_{\lambda=1}^2\sum_{\bf k} g_{\bf{k},\lambda}^b e^{i(\omega_{bc}-\nu_k)t}|b\rangle\langle c|\,\hat{a}_{\bf{k},\lambda} + {\rm h.c.},
\end{equation}
where $g_{\bf{k},\lambda}^a, g_{\bf{k},\lambda}^b$ [rad/s] are the coupling terms between the k-th mode of the reservoir (with wave vector $\bf{k}$, polarization $\lambda$ and angular frequency $\nu_k$) and transitions $|a\rangle\leftrightarrow|c\rangle,|b\rangle\leftrightarrow|c\rangle$. Given that 
\begin{equation}\label{eq:def1}
g_{\bf{k},\lambda}^r \equiv -\frac{\bm{\mu}_{rc}\cdot\bm{\epsilon}_{\bf{k},\lambda}}{\hbar}\sqrt{\frac{\hbar\nu_k}{2\varepsilon_0V}}    
\end{equation}
with $r=a,b$, the coupling terms depend on the quantization volume $V$ (i.e., the spatial region where the radiation field effectively interact with the system), on the electric dipole moment matrix element $\bm{\mu}_{rc}=\langle r|\bm{\mu}|c\rangle$, relative to the $|r\rangle\leftrightarrow|c\rangle$ optical transition, and on the unitary polarization vector $\bm{\epsilon}_{\bf{k},\lambda}$ of the radiation. Both $\bm{\mu}_{rc}$ and $\bm{\epsilon}_{\bf{k},\lambda}$ are assumed real. Then, $\hat{a}_{\bf{k},\lambda}$ and its Hermitian transpose $\hat{a}_{\bf{k},\lambda}^\dag$ are the annihilation and creation operators of the bosonic field, respectively. Also note that the definition (\ref{eq:def1}) originates from the \textit{dipole approximation}, where the spatial dependence of the field is ignored.

The time evolution of the whole system is governed by the Liouville-von Neumann differential equation for $\hat{\rho}(t)$: 
\begin{equation}\label{eq:one}
\frac{d\hat{\rho}(t)}{dt} = -\frac{i}{\hbar}[H_I, \hat{\rho}(t)]\,.
\end{equation}
By integrating \eqref{eq:one} and inserting the result back into (\ref{eq:one}), the following integro-differential form of the Liouville-von Neumann equation is obtained:
\begin{equation}
\frac{d\hat{\rho}(t)}{dt} = -\frac{i}{\hbar}\left[ \hat{H}_I(t), \hat{\rho}(0)
\right] - \frac{1}{\hbar^2}\int_{0}^{t}\Big[ \hat{H}_I(t), \left[ \hat{H}_I(t'), \hat{\rho}(t')\right] \Big] dt'.
\label{eq:two}
\end{equation}

In several real-life scenarios, the three-level system and the reservoir are weakly coupled, implying that the influence of the system on the reservoir is negligible. Consequently, assuming that the initial state of the total system is the separable state $\hat{\rho}(0) = \hat{\rho}_S(0) \otimes \hat{\rho}_R(0)$, the state $\hat{\rho}(t)$ at any given time $t$ can be approximated by the tensor product $\hat{\rho}(t) \approx \hat{\rho}_S(t) \otimes \hat{\rho}_R(0)$. This assumption is known as the \textit{Born} or \textit{weak coupling approximation}.
Therefore, using the partial trace over the reservoir degrees of freedom, the reduced state for the quantum system only is
\begin{equation}\label{eq:four}
\frac{d\hat{\rho}_S(t)}{dt} = -\frac{i}{\hbar}{\rm Tr}_B\left[ \hat{H}_I(t), \hat{\rho}_S(0) \otimes \hat{\rho}_R(0) \right] - \frac{1}{\hbar^2}\int_0^t {\rm Tr}_B \Big[ \hat{H}_I(t), \left[ \hat{H}_I(t'), \hat{\rho}_S(t') \otimes \hat{\rho}_B(0)\right]
\Big] dt'.
\end{equation}

Let us now analyze the first term on the right-hand side of \eqref{eq:four}, which is associated with the {\it coherent} part of the dynamics and we now denote as $\frac{d\hat{\rho}_S(t)}{dt}|_{\text{coh}}$. In doing this, let us insert $\hat{H}_I(t)$ in (\ref{eq:four}), and upon further calculation for this coherent term we get that
\begin{equation}\label{eq:rhoS_dot_coh}
\left.\frac{d\hat{\rho}_S(t)}{dt}\right|_{\text{coh}} = 
-i\sum_{\lambda=1}^2\sum_{\bf{k}}g_{\bf{k},\lambda}^ae^{i(\omega_{ac}-\nu_k)t}\langle\hat{a}_k\rangle[\hat{\sigma}_{ac}^+,\hat{\rho}_S(0)] -i\sum_{\lambda=1}^2\sum_{\bf{k}} g_{\bf{k},\lambda}^be^{i(\omega_{bc}-\nu_k)t}\langle\hat{a}_k\rangle[\hat{\sigma}_{bc}^+,\hat{\rho}_S(0)] + {\rm h.c.}\,, 
\end{equation}
where we use the transition operators $\hat{\sigma}_{ac}^{+} \equiv |a\rangle\langle c|, \hat{\sigma}_{bc}^{+} \equiv |b\rangle\langle c|$ and their Hermitian transpose $\hat{\sigma}_{ac}^-$ and $\hat{\sigma}_{bc}^-$. Moreover, \eqref{eq:rhoS_dot_coh} contains also the expectation values $\langle \hat{a}_{\bf{k},\lambda}\rangle$ and $\langle \hat{a}_{\bf{k},\lambda}^{\dagger}\rangle$ that are computed with respect to the composite state of the reservoir. Here, we assume that the modes of the reservoir are distributed among a mixture of uncorrelated thermal equilibrium states at temperature $T$. In this way, the expectation value and the correlation function of the reservoir's operators, computed with respect to the mixture of local thermal states for each mode, get the following values:
\begin{eqnarray}
    \langle \hat{a}_{\bf{k},\lambda} \rangle &=& \langle \hat{a}_{\bf{k},\lambda}^\dag \rangle = 0\label{eq:thermal_occup_number} \\
    \langle \hat{a}_{\bf{k},\lambda}^\dag \hat{a}_{\bf{k'},\lambda} \rangle &=& \bar{n}_k\delta_{\bf{k},\bf{k'}}\delta_{\lambda,\lambda'}\label{eq:five} \\
    \langle \hat{a}_{\bf{k},\lambda} \hat{a}_{\bf{k'},\lambda}^\dag \rangle &=& (\bar{n}_k+1)\delta_{\bf{k},\bf{k'}}\delta_{\lambda,\lambda'}\label{eq:six} \\
    \langle \hat{a}_{\bf{k},\lambda} \hat{a}_{\bf{k'},\lambda} \rangle &=& \langle \hat{a}_{\bf{k},\lambda}^\dag \hat{a}_{\bf{k'},\lambda}^\dag \rangle = 0\label{eq:seven}
\end{eqnarray}
where $\bar{n}_k \equiv [\exp(\hbar\nu_k/(k_B T))-1]^{-1}$ is the average occupation number of the k-th thermal mode of the incoherent field, with $k_B$ the Boltzmann constant and $\delta$ the Dirac delta function. Thus, by substituting \eqref{eq:thermal_occup_number} in \eqref{eq:rhoS_dot_coh}, we end-up to
\begin{equation}
\left.\frac{d\hat{\rho}_S(t)}{dt}\right|_{\text{coh}}  = 0 \,.
\end{equation}

We now analyze the second term on the right-hand side of Eq.~(\ref{eq:four}), which is associated with the {\it incoherent} part of the dynamics. In order to attain the simplified expression of the latter, we need to expand the double commutator in Eq.~(\ref{eq:four}) after substituting Eqs.~\eqref{eq:interaction_H}-(\ref{eq:def1}). 
The final expression we now provide (the complete derivation is in Methods, Sec.~\ref{appendix_model}) comes from applying the {\it Weisskopf-Wigner approximation} that assumes the all the frequency modes $\nu_k$ of the radiation field are closely spaced within a spherical volume. We also consider that the frequency modes are approximately constant around $\omega_{ac} \simeq \omega_{bc}$ for any $k$, given the presence of exponential terms of the form $e^{\pm i(\omega_{rc}-\nu_k)(t-t')}$ ($r=a,b$) that, oscillating rapidly with respect to the energy scale of the system dynamics, are negligible. At the end of the computation, the double commutator in Eq.~(\ref{eq:four}) comprises terms of the form 
\begin{equation}\label{eq:thirteen_main}
\frac{\omega_{rc}^3\left|\bm{\mu}_{ac}\right|^2}{\hbar6\pi^2\varepsilon_0c^3} \int_0^t \Big[\bar{n}\left(\hat{\sigma}_{ac}^-\hat{\sigma}_{ac}^+\hat{\rho}_S(t')- \hat{\sigma}_{ac}^+\hat{\rho}_S(t')\hat{\sigma}_{ac}^-\right) +(\bar{n}+1)\left(\hat{\rho}_S(t')\hat{\sigma}_{ac}^+\hat{\sigma}_{ac}^{-} - \hat{\sigma}_{ac}^-\hat{\rho}_S(t')\hat{\sigma}_{ac}^+\right)\Big]e^{-i\omega_{rc}(t-t')}2\pi\delta(t-t')dt' 
\end{equation}
and 
\begin{equation}\label{eq:fourteen_main}
\frac{\omega_{ac}^3\left(\bm{\mu}_{ac}\cdot\bm{\mu}_{bc}\right)}{\hbar6\pi^2\varepsilon_0c^3}\int_0^t 
\Big[ \bar{n}\left(\hat{\rho}_S(t')\hat{\sigma}_{bc}^-\hat{\sigma}_{ac}^+ 
 -\hat{\sigma}_{ac}^+\hat{\rho}_S(t')\hat{\sigma}_{bc}^-\right) + (\bar{n}+1)\left(\hat{\sigma}_{ac}^+\hat{\sigma}_{bc}^-\hat{\rho}_S(t') - \hat{\sigma}_{bc}^-\hat{\rho}_S(t')\hat{\sigma}_{ac}^+\right) \Big] e^{i\omega_{ac}(t-t')}2\pi\delta(t-t')dt',
\end{equation}
where $\bar{n} \equiv [\exp(\hbar\omega_{ac}/k_BT)-1]^{-1}$. We emphasize that Eqs.~\eqref{eq:thirteen_main}-\eqref{eq:fourteen_main} are derived under the assumption of \textit{isotropic} and \textit{unpolarized} radiation. Additionally, they have clearly non-Markovian traits leading to memory effects in the dynamics of the quantum system, given the dependence of the right-hand-side of the equations to all the ``history'' of $\hat{\rho}_S(t')$ from $0$ to $t$. Hence, to get a quantum Markovian master equation, we apply \textit{Markov approximation} that is valid whenever the correlations between the quantum system and the reservoir decay rapidly in comparison with the rate of change of the system's state. Therefore, under the Markov approximation, the master equation governing the system dynamics only depends on $\hat{\rho}_S(t)$ at time $t$ and not on its past history. Hence, setting the upper limit of the integral to $\infty$ and $\hat{\rho}_S(t')=\hat{\rho}_S(t)$, assuming $t=t'$ $\forall t$, Eq.~(\ref{eq:four}) reads as
\begin{eqnarray}\label{eq:sixteen}
&&\frac{d\hat{\rho}_S(t)}{dt} = \left.\frac{d\hat{\rho}_S(t)}{dt}\right|_{\text{incoh}} =\nonumber \\ 
&& - \frac{\gamma_a}{2}\Big[ \bar{n}\big(\hat{\sigma}_{ac}^-\hat{\sigma}_{ac}^+\hat{\rho}_S(t)-\hat{\sigma}_{ac}^+\hat{\rho}_S(t)\hat{\sigma}_{ac}^-\big)+(\bar{n}+1)\big(\hat{\rho}_S(t)\hat{\sigma}_{ac}^+\hat{\sigma}_{ac}^- + 
 - \hat{\sigma}_{ac}^-\hat{\rho}_S(t)\hat{\sigma}_{ac}^+\big) \Big] \nonumber \\
&& - \frac{\gamma_b}{2}\Big[ \bar{n}\big(\hat{\sigma}_{bc}^-\hat{\sigma}_{bc}^+\hat{\rho}_S(t)-\hat{\sigma}_{bc}^+\hat{\rho}_S(t)\hat{\sigma}_{bc}^-\big)+(\bar{n}+1)\big(\hat{\rho}_S(t)\hat{\sigma}_{bc}^+\hat{\sigma}_{bc}^- - \hat{\sigma}_{bc}^-\hat{\rho}_S(t)\hat{\sigma}_{bc}^+\big) \Big] \nonumber \\ 
&& - p\frac{\sqrt{\gamma_a\gamma_b}}{2}\Big[ \bar{n}\big(\hat{\rho}_S(t)\hat{\sigma}_{bc}^-\hat{\sigma}_{ac}^+ - \hat{\sigma}_{ac}^+\hat{\rho}_S(t)\hat{\sigma}_{bc}^- \big)+(\bar{n}+1)\big(\hat{\sigma}_{ac}^+\hat{\sigma}_{bc}^-\hat{\rho}_S(t) - \hat{\sigma}_{bc}^-\hat{\rho}_S(t)\hat{\sigma}_{ac}^+\big) \Big] \nonumber \\
&& - p\frac{\sqrt{\gamma_a\gamma_b}}{2}\Big[ \bar{n}\big(\hat{\rho}_S(t)\hat{\sigma}_{ac}^-\hat{\sigma}_{bc}^+ - \hat{\sigma}_{bc}^+\hat{\rho}_S(t)\hat{\sigma}_{ac}^- \big)+(\bar{n}+1)\big(\hat{\sigma}_{bc}^+\hat{\sigma}_{ac}^-\hat{\rho}_S(t) - \hat{\sigma}_{ac}^-\hat{\rho}_S(t)\hat{\sigma}_{bc}^+\big) \Big] + \rm{h.c.}\,.
\end{eqnarray}
In Eq.~\eqref{eq:sixteen},
\begin{eqnarray}
    \gamma_r &\equiv& \frac{\omega_{rc}^3\left|\mu_{rc}\right|^2}{\hbar3\pi \varepsilon_0 c^3} \label{eq:def2} \\
    p &\equiv& \frac{\bm{\mu}_{ac}\cdot\bm{\mu}_{bc}}{\left|\bm{\mu}_{ac}\right|\left|\bm{\mu}_{bc}\right|}=\cos\Theta \label{eq:def3}
\end{eqnarray}
where $\gamma_r$ denotes the \textit{spontaneous decay rate} from level $|r\rangle$ to the ground level $|c\rangle$, and $p$ is the \textit{alignment parameter} between the transition dipole moments of the transitions $|a\rangle\leftrightarrow|c\rangle,|b\rangle\leftrightarrow|c\rangle$. Thus, $\Theta$ is the angle between the two electric dipole moments.

So far, we have determined a quantum Markovian master equation [Eq.~(\ref{eq:sixteen})] in the interaction picture that describes the evolution of a V-type three-level system in the presence of isotropic, unpolarized, broadband radiation. There, the terms $p\sqrt{\gamma_a\gamma_b}$ model the quantum interference that arises from incoherent pumping and spontaneous emission processes~\cite{Gong}. Notably, when the matrix elements $\bm{\mu}_{ac}, \bm{\mu}_{bc} $ are orthogonal, thus resulting in $p = 0$, such an interference is absent. The magnitude of the quantum interference is maximized when the transition dipole moments are either parallel ($p=1$) or anti-parallel ($p=-1$).

Finally, we obtain the equation of motion for $\hat{\rho}_S(t)$ in the Schr{\"o}dinger picture by adding the Hamiltonian $\hat{H}_S$ of the three-level system in the coherent part of the differential equation of $\hat{\rho}_S(t)$. Formally, it entails to solve the differential equation
\begin{equation}
\frac{d\hat{\rho}_S(t)}{dt} = -\frac{i}{\hbar}{\rm Tr}_B\left[ \hat{H}_I(t) + \hat{H}_S \otimes \hat{I}_B, \hat{\rho}_S(0) \otimes \hat{\rho}_B(0) \right] 
- \frac{1}{\hbar^2}\int_0^t {\rm Tr}_B\left[ \hat{H}_I(t),\left[\hat{H}_I(t'),\hat{\rho}_S(t')\otimes\hat{\rho}_B(0)\right]\right]dt'
\end{equation}
with $\hat{I}_B$ denoting the identity operator in the Hilbert space of the reservoir. Hence, by including the explicit expression of $\hat{H}_S = \sum_k E_k |k\rangle\!\langle k|$ in the differential equation \eqref{eq:sixteen} and decomposing $\hat{\rho}_S(t)$ in its elements $\langle k|\hat{\rho}_S(t)|j\rangle \equiv \rho_{kj}(t)$ with $k,j=a,b,c$, the following set of differential equations for each $\rho_{kj}(t)$ can be obtained: 
\begin{equation}\label{eq:seventeen}
\begin{cases}
\displaystyle{ \frac{ d\rho_{aa}(t) }{dt} = -\gamma_a\left(\bar{n} +1\right)\rho_{aa}(t) + \gamma_a \bar{n} \rho_{cc}(t) -p\sqrt{\gamma_a \gamma_b }\left(\bar{n} +1\right)\rm{Re}[\rho_{ab}(t)] }\\
\displaystyle{ \frac{ d\rho_{bb}(t) }{dt} = -\gamma_b\left(\bar{n} +1\right)\rho_{bb}(t) + \gamma_b \bar{n} \rho_{cc}(t) -p\sqrt{\gamma_a \gamma_b }\left(\bar{n} +1\right)\rm{Re}[\rho_{ab}(t)] }\\
\displaystyle{ \frac{ d\rho_{cc}(t) }{dt} = -\left(\gamma_a+\gamma_b\right)\bar{n}\rho_{cc}(t) + \left( \bar{n} +1 \right)\left( \gamma_a\rho_{aa}(t) + \gamma_b \rho_{bb}(t) \right) + 2p\sqrt{\gamma_a \gamma_b }\left( \bar{n} +1 \right)\rm{Re}[\rho_{ab}(t)] }\\
\displaystyle{ \frac{ d\rho_{ab}(t) }{dt} = -p\frac{\sqrt{\gamma_a \gamma_b }}{2}\left(\bar{n} +1\right)\left(\rho_{aa}(t)+\rho_{bb}(t)\right) + p\sqrt{\gamma_a \gamma_b }\bar{n}\rho_{cc}(t) -\left[\frac{\gamma_a+\gamma_b}{2}\left(\bar{n} +1\right)+i\rm{\Delta}\right]\rho_{ab}(t) }
\end{cases}
\end{equation}
together with
\begin{equation}\label{eq:eighteen}
\begin{cases}
\displaystyle{ \frac{ d\rho_{ac}(t) }{dt} = -p\frac{\sqrt{\gamma_a \gamma_b }}{2}\left(\bar{n} +1\right)\rho_{bc}(t) -\left[\frac{\gamma_b}{2}\bar{n}+\frac{\gamma_a}{2}\left(2\bar{n}+1\right) + i\frac{\omega_{ac}}{2}\right]\rho_{ac}(t) } \\
\displaystyle{ \frac{ d\rho_{bc}(t) }{dt} = -p\frac{\sqrt{\gamma_a \gamma_b }}{2}\left(\bar{n} +1\right)\rho_{ac}(t) -\left[\frac{\gamma_a}{2}\bar{n}+\frac{\gamma_b}{2}\left(2\bar{n}+1\right) + i\frac{\omega_{bc}}{2}\right]\rho_{bc}(t) \,,} 
\end{cases}
\end{equation}
where the incoherent pumping rates $R_r \equiv \bar{n}\gamma_r$ of the transitions $|r\rangle\leftrightarrow|c\rangle$ $(r=a,b)$ are associated with the absorption and stimulated-emission processes due to the incoherent light source. Note that, if $p=0$, then Eqs.~\eqref{eq:seventeen}-\eqref{eq:eighteen} for the quantum system dynamics simplify to the standard Pauli rate equations~\cite{Koyu:Tscherbul,Dodin:Tscherbul}.

Eqs.~(\ref{eq:seventeen}) and (\ref{eq:eighteen}) correspond to two independent sub-processes of the quantum system's evolution~\cite{Agarwal:Menon}. Thus, the solution to the differential equations therein contained can be achieved by solving two distinct systems of {\it linear} equations, \emph{i.e.}, 
\begin{equation}\label{eq:nineteen}
\frac{ d{\bf{x}}(t) }{dt} = A{\bf x}(t) \quad \text{and} \quad
\frac{ d{\bf{z}}(t) }{dt} = C{\bf z}(t)
\end{equation}
with state vectors
\begin{eqnarray}
{\bf x}(t) &\equiv& \Big( \rho_{aa}(t), \rho_{bb}(t), \rho_{cc}(t), {\rm Re}[\rho_{ab}(t)], {\rm Im}[\rho_{ab}(t)]\Big)^T \\
{\bf z}(t) &\equiv& \Big( {\rm Re}[\rho_{ac}(t)], {\rm Im}[\rho_{ac}(t)], {\rm Re}[\rho_{bc}(t)], {\rm Im}[\rho_{bc}(t)]\Big)^T \,.
\end{eqnarray}
Note that, differently from previous approaches~\cite{Koyu:Tscherbul,Dodin:Brumer,Dodin:Brumer2}, the vector ${\bf x}$ includes the population of the ground level $\rho_{cc}(t)$ rather than imposing the constraint $\rho_{cc}(t) = 1 - \rho_{aa}(t) - \rho_{bb}(t)$. This choice is needed to preserve the unity of the density operator $\hat{\rho}_S(t)$ at any time $t$ when passing from both the linear differential equations in~\eqref{eq:nineteen}. In Eq.~\eqref{eq:nineteen}, the matrices $ A,C $ of coefficients are equal to
\begin{eqnarray}
A &=& 
\begin{pmatrix}
-\gamma_{a}\left(\bar{n}+1\right) & 0 & \gamma_{a}\bar{n} & \,\, -p\sqrt{\gamma_a\gamma_b}\left(\bar{n}+1\right) & \,\,\,\,0 \\
0 & -\gamma_{b}\left(\bar{n}+1\right)  & \gamma_{b}\bar{n} & \,\, -p\sqrt{\gamma_a\gamma_b}\left(\bar{n}+1\right) & \,\,\,\,0  \\
\gamma_{a}\left(\bar{n}+1\right) & \gamma_{b}\left(\bar{n}+1\right) & - \left(\gamma_{a}+\gamma_{b}\right)\bar{n} & \,\, 2p\sqrt{\gamma_a\gamma_b}\left(\bar{n}+1\right) & \,\,\,\,0  \\
-\frac{p}{2}\sqrt{\gamma_a\gamma_b}\left(\bar{n}+1\right) & -\frac{p}{2}\sqrt{\gamma_a\gamma_b}\left(\bar{n}+1\right) & p\sqrt{\gamma_a\gamma_b}\bar{n} & \,\, -\frac{\gamma_a+\gamma_b}{2}\left(\bar{n}+1\right) & \,\,\,\, {\rm\Delta}\\
0 & 0 & 0 & \,\, - {\rm\Delta} &  -\frac{\gamma_a+\gamma_b}{2}\left(\bar{n}+1\right)
\end{pmatrix}
\\
C &=&
\begin{pmatrix}
- \left[\bar{n}\left(\gamma_a+\frac{\gamma_b}{2}\right)+\frac{\gamma_a}{2}\right] &  \omega_{ac} & -\frac{p}{2}\sqrt{\gamma_a\gamma_b}\left(\bar{n}+1\right) & 0 \\
-\omega_{ac} & - \left[\bar{n}\left(\gamma_a+\frac{\gamma_b}{2}\right)+\frac{\gamma_a}{2}\right] & 0 & -\frac{p}{2}\sqrt{\gamma_a\gamma_b}\left(\bar{n}+1\right)\\
-\frac{p}{2}\sqrt{\gamma_a\gamma_b}\left(\bar{n}+1\right) & 0 & - \left[\bar{n}\left(\gamma_b+\frac{\gamma_a}{2}\right)+\frac{\gamma_b}{2}\right] & \omega_{bc}\\
0 & -\frac{p}{2}\sqrt{\gamma_a\gamma_b}\left(\bar{n}+1\right) & -\omega_{bc} &  - \left[\bar{n}\left(\gamma_b+\frac{\gamma_a}{2}\right)+\frac{\gamma_b}{2}\right]
\end{pmatrix}
\end{eqnarray}
We numerically solve the homogeneous differential equations (\ref{eq:nineteen}) via exponentiation, namely
\begin{eqnarray}
    {\bf x}(t) &=& e^{At}{\bf x}(0)\label{eq:sol_x_t} \\
    {\bf z}(t) &=& e^{Ct}{\bf z}(0)\label{eq:sol_z_t}
\end{eqnarray}
with ${\bf x}(0),\,{\bf z}(0)$ denoting the initial states in this representation. The exponential of the matrices $A,C$ is computed using the \textsc{Matlab} function \texttt{expm}, which employs the scaling and squaring algorithm of Higham~\cite{Higham}. 

Analytical solutions of Eqs.~\eqref{eq:seventeen} and~\eqref{eq:eighteen} have been demonstrated in previous studies~\cite{Koyu:Tscherbul,Dodin:Brumer}. These solutions exhibit different behaviors depending on the value of the ratio ${\rm\Delta}/\bar{\gamma}$ (between the energy splitting ${\rm\Delta}$ among the excited states and the average decay rate $\bar{\gamma}$), as well as on the average photon number $\bar{n}$ and on the alignment parameter $p$. Specifically, three regimes emerge: the overdamped, the underdamped, and the critical regimes. It is noteworthy that only in the overdamped regime quasi-stationary Fano coherences can be established, thus resulting in a prolonged coherence lifetime. Under the \textit{weak pumping} condition ($\bar{n}<1$) with $p<|1|$, achieving the overdamped regime is possible when ${\rm \Delta}/\bar{\gamma}\ll 1$. However, under the \textit{strong pumping} condition ($\bar{n}>1$), the requirement ${\rm \Delta}/\bar{\gamma}\ll 1$ can be relaxed, which means a value of ${\rm\Delta}$ much larger than $\bar{\gamma}$ without compromising the quasi-stationarity of coherences. This rationale will guide our selection of ${\rm \Delta}/\bar{\gamma}\ll 1$ in the analyses below.

\subsection{Quasiprobabilities}\label{sec_quasiprobs}

In the previous section, we have introduced a quantum Markovian master equation that originates Fano quantum coherences. In this regard, we recall that they can arise by illuminating a quantum system with an incoherent source, provided the system has a discrete number of levels and some of these levels are almost-degenerate. It is not possible to generate Fano coherences on a two-level system (a qubit), but it becomes possible on a three-level system admitting two almost-degenerate energy levels, as we are going to show below with a detailed analysis.

Since our aim is to determine the energy cost in generating Fano coherences and then to understand the role of energy fluctuations beyond the average, we introduce the {\it Kirkwood-Dirac quasiprobabilities} (KDQ)~\cite{yunger2018quasiprobability,lupu2021negative,DeBievrePRL2021,LostaglioKirkwood2022,SantiniPRB2023,BudiyonoPRAquantifying,wagner2023quantum}. Thanks to the latter, we can describe the two-time statistics of the energy outcomes originated from evaluating the Hamiltonian of the quantum system in two distinct times.

Let us thus formalize the physical context we will work in, as well as the definition of the KDQ. We will consider a three-level system with time-independent Hamiltonian $\hat{H}_S = \sum_{k=1}^{3}E_{k}\hat{\Pi}_{k}$, with $E_k$ denoting the energies of the system (eigenvalue of $\hat{H}_S$) and $\hat{\Pi}_k \equiv |E_k\rangle\!\langle E_k|$ the corresponding projectors ($|E_k\rangle$ are the eigenstates of $\hat{H}_S$). The three-level system is initialized in the initial density operator $\hat{\rho}_S(0)$ and then is subjected to the open quantum map $\Phi[\cdot]$ that returns the density operator of the system at time $t$, \emph{i.e.}, $\hat{\rho}_S(t) = \Phi[\hat{\rho}_S(0)]$.
It is also responsible for the generation of Fano coherences under specific conditions; in this regard, we will show practical examples below.
Hence, the KDQ describing the statistics of the energy-changes in the interval $[t_1,t_2]$ is defined as
\begin{equation}\label{eq:def_KDQ}
    q_{\ell,j} = {\rm Tr}\left[ \hat{\Pi}_{j}\,\Phi\left[\hat{\Pi}_{\ell}\,\hat{\rho}_S(0)\right] \right],
\end{equation}
where $\hat{\Pi}_j$ and $\hat{\Pi}_{\ell}$ are the $j$-th and $\ell$-th projectors of $\hat{H}_S$ evaluated at times $t_1$ and $t_2$ respectively. Each quasiprobability $q_{\ell,j}$ is associated to the $(\ell,j)$-th realization ${\rm\Delta}E_{\ell,j} \equiv E_{j} - E_{\ell}$ of the energy-change ${\rm\Delta}E$, which is given by the difference of the system energies evaluated at times $t_1$ and $t_2$. We recall that the real parts of KDQ are also known as Margenau-Hill quasiprobabilities (MHQ)~\cite{HalliwellPRA2016,levy2020quasiprobability,hernandez2022experimental,PeiPRE2023}, and has recently found several applications in quantum thermodynamics.

It is worth providing some properties of KDQ~\cite{LostaglioKirkwood2022} in the case-study we are here analyzing:
\begin{itemize}
\item[(i)] 
The sum of KDQ is equal to $1$: $\sum_{\ell,j}q_{\ell,j}=1$.
\item[(ii)] 
The {\it unperturbed} marginals are obtained:
\begin{eqnarray}
    \sum_{\ell}q_{\ell,j} &=& p_{j}(t_2) \equiv {\rm Tr}\left[ \hat{\Pi}_{j}\Phi[\hat{\rho}_S(0)] \right] = {\rm Tr}\left[ \hat{\Pi}_{j}\hat{\rho}_{S}(t) \right] \\
    \sum_{j}q_{\ell, j} &=& p_{\ell}(t_1) \equiv {\rm Tr}\left[ \hat{\Pi}_{\ell}\hat{\rho}_S(0) \right],
\end{eqnarray}
where ``unperturbed'' means that the marginals are equal to the probabilities to measure the system at the single times $t_1$ and $t_2$ respectively, as given by the Born rule.   
Let us observe that, if $[\hat{\rho}_S(0),\hat{H}_S] \neq 0$ for some $\hat{\rho}_S(0)$ and $\hat{H}_S$, then the unperturbed marginal $p_{j}(t_2)$ at time $t_2$ is not obtained by the two-point measurement (TPM) scheme~\cite{campisi2011colloquium}. The latter, indeed, cancels the off-diagonal terms of $\hat{\rho}_S(0)$ with respect to the eigenbasis of $\hat{H}_S$, due to the initial projective measurement at time $t_1$ whose effect is to induce decoherence.
\item[(iii)] 
The KDQ are linear in the initial density operator $\hat{\rho}_S(0)$. This means that, given any admissible decomposition of $\hat{\rho}_S(0)$ [say $\hat{\rho}_S(0) = \hat{\rho}_S^{(1)}(0) + \hat{\rho}_S^{(2)}(0)$], \eqref{eq:def_KDQ} splits in two terms, one linearly dependent on $\hat{\rho}_S^{(1)}(0)$ and the other on $\hat{\rho}_S^{(2)}(0)$, \emph{i.e.},
\begin{equation}\label{eq:linear_decomposition_KDQ}
    q_{\ell, j} = q_{\ell, j}^{(1)} + q_{\ell, j}^{(2)} 
\end{equation}
with $q_{\ell, j}^{(n)} = {\rm Tr}[ \hat{\Pi}_{\ell}\,\Phi[\hat{\Pi}_{j}\hat{\rho}_S^{(n)}(0)]]$, $n=1,2$. A choice that is commonly adopted is to take $\hat{\rho}_S^{(1)}(0)$ as the matrix that solely contains the diagonal terms of $\hat{\rho}_S(0)$, and $\hat{\rho}_S^{(2)}(0)$ as the matrix comprising only the off-diagonal terms.
\item[(iv)]
Under the commutative condition $[\hat{\rho}_S(0),\hat{H}_S] = 0$, the KDQ are equal to the joint probabilities 
\begin{equation}
p_{\ell, j}^{\rm TPM} \equiv {\rm Tr}\left[ 
\hat{\Pi}_{j}\Phi\left[\hat{\Pi}_{\ell}\hat{\rho}_S(0)\hat{\Pi}_{\ell}\right]
\right]     
\end{equation}
returned by the TPM scheme.
\item[(v)]
KDQ are in general complex numbers and can lose positivity, \emph{i.e.}, they can admit negative real parts and imaginary parts different from zero. In fact, as prescribed by the no-go theorems in Refs.~\cite{PerarnauLlobetPRL2017,LostaglioKirkwood2022}, we recall that the non-positivity of KDQ is due to asking for unperturbed marginals and (quasi)joint probabilities of the distribution of ${\rm\Delta}E$ that are linear in the initial density operator $\hat{\rho}_S(0)$, whenever $[\hat{\rho}_S(0),\hat{H}_S] \neq 0$.
The presence of non-positivity is a proof of quantum contextuality~\cite{SpekkensPRL2008,PuseyPRL2014,KunjwalPRA2019}, and denotes a form of non-classicality for the two-time statistics of energy outcomes. The non-positivity of KDQ can be quantified by the {\it non-positivity functional}~\cite{AlonsoPRL2019,ArvidssonShukurJPA2021,hernandez2022experimental,LostaglioKirkwood2022} 
\begin{equation}\label{eq:def_aleph}
    \aleph \equiv - 1 + \sum_{\ell, j}\big|q_{\ell, j}\big|
\end{equation}
that is equal to $1$ when all the quasiprobabilities are positive real numbers.
\end{itemize}

\subsection{Quantumness certification}
\label{sec_certification}

In this section we certify that the generation of Fano coherences can be accompanied by a distribution of quasiprobabilities for the energy change of a V-type three-level system, which exhibits negativity (albeit with no imaginary parts). The presence of the latter results from initializing the three-level system in a superposition of the wave-functions comprising the energy eigenbasis; this means that in such a basis quantum coherences have to be included. This occurs for specific parameter settings that we will analyze in more details in Sec.~\ref{sec_thermodynamics}. Interestingly, there is also a subset of parameters' values such that solely the quantum coherence in the initial state of the system (leading to negativity) is responsible for an amount of extractable work 
\begin{equation}
-\langle{\rm\Delta}E(t)\rangle = -\sum_{\ell,j}q_{\ell,j}{\rm\Delta E}_{\ell,j}(t) = {\rm Tr}\left[ \hat{H}_S \left( \hat{\rho}_S(0) - \hat{\rho}_S(t) \right) \right]    
\end{equation}
larger than zero for any time $t$, with $\hat{H}_S$ time-independent.

\begin{figure}[t]
\centering
\includegraphics[width=0.6\textwidth]{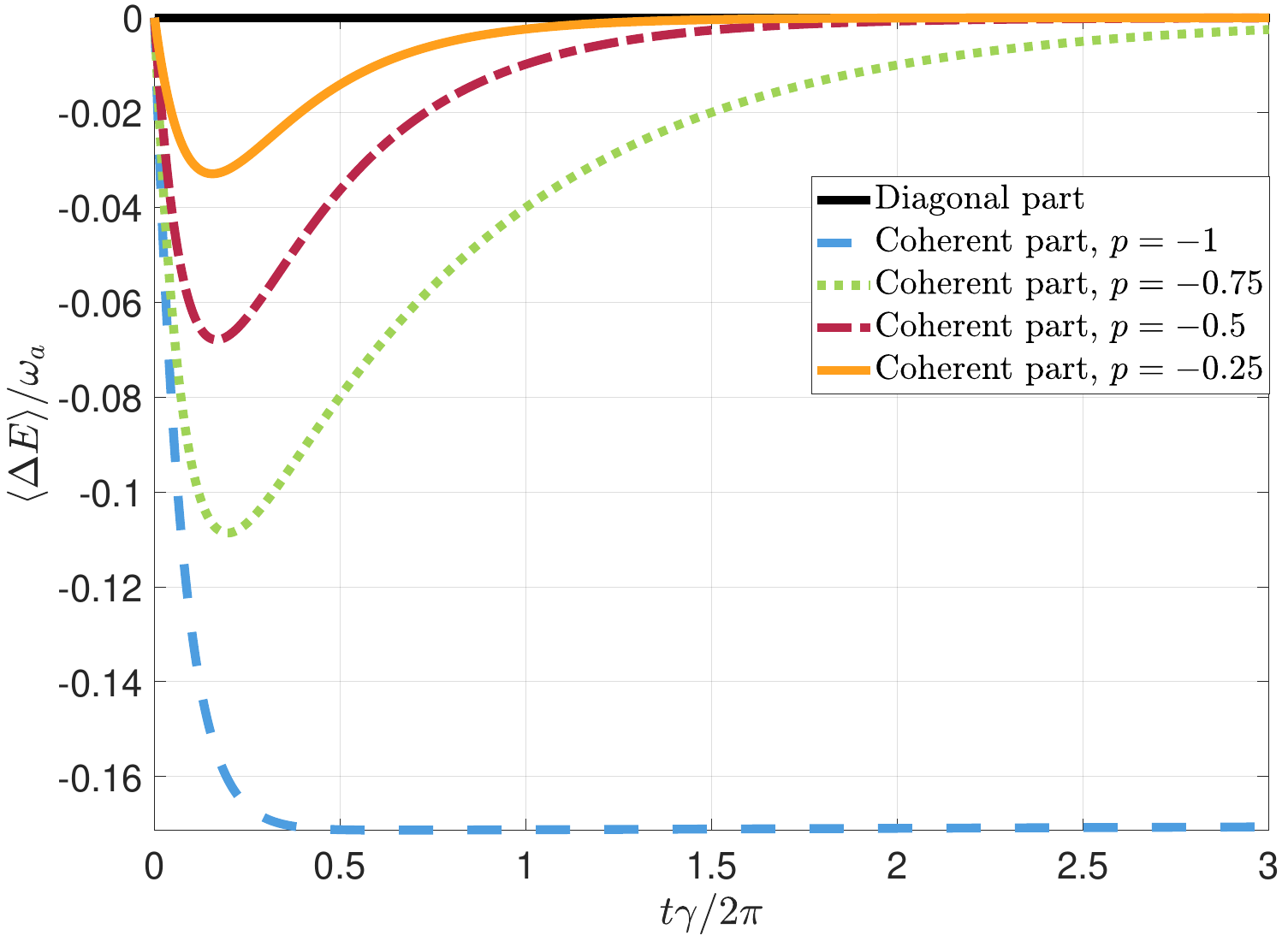}
\caption{
Average energy-change $\langle {\rm\Delta} E\rangle$, re-scaled by $\omega_a$, as a function of the dimensionless time $t\gamma/(2\pi)$, which we obtain by numerically computing the corresponding KDQ distribution. The dynamics of the three-level system subjected to an incoherent light source, entering in the quasiprobabilities, is provided by Eqs.~\eqref{eq:nineteen}. The black solid line denotes the contribution $\langle{\rm \Delta}E\rangle_{\rm diag}$ of the average energy-change that corresponds solely to the diagonal elements, contained in ${\rm diag}(\hat{\rho}_S(0))$, of the initial state $\hat{\rho}_S(0)$. It can be verified that $\langle{\rm \Delta}E\rangle_{\rm diag}$ is equal to zero for any value of $p$. On the other hand, all the other curves in the figure refer to the contribution $\langle{\rm \Delta}E\rangle_{\rm coh}$ of the average energy-change depending on $\chi_S$, matrix containing the off-diagonal elements of $\hat{\rho}_S(0)$, for $p=-0.25,-0.5,-0.75,-1$.  
}
\label{fig:extractable_work}
\end{figure}
\begin{figure}[h]
\centering
\includegraphics[width=0.8\textwidth]{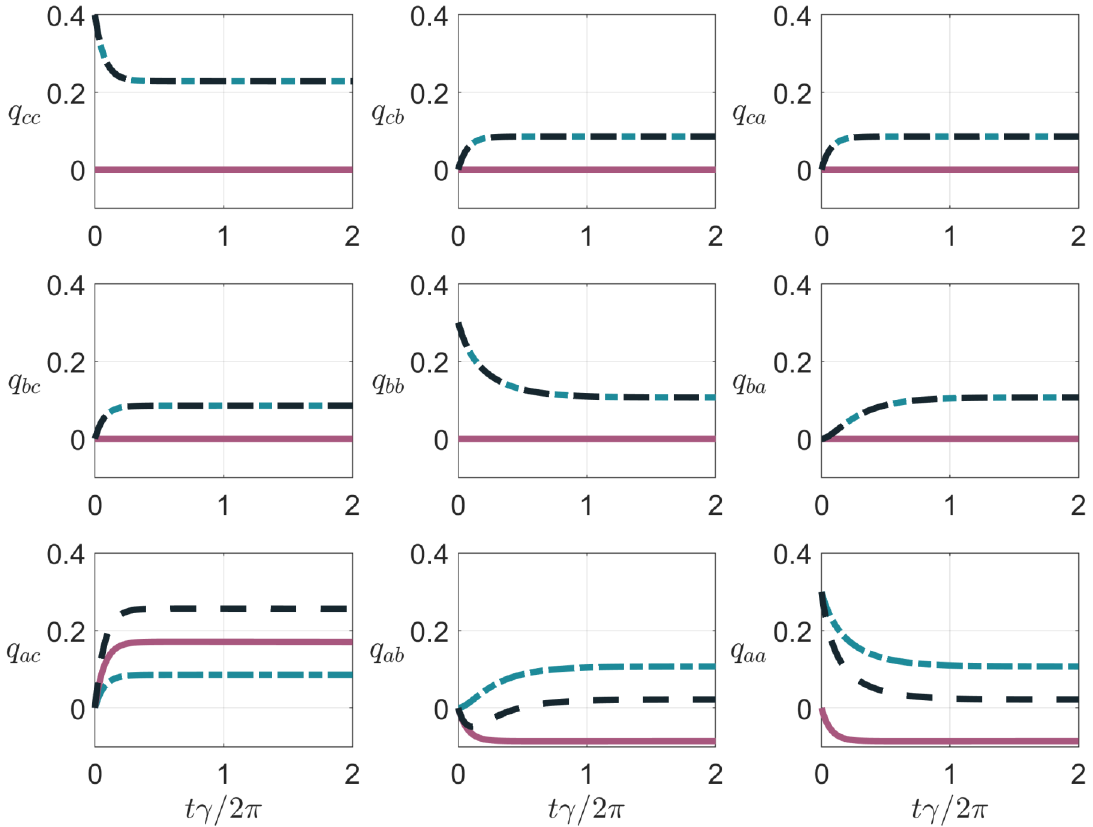}
\caption{
Kirkwood-Dirac quasiprobabilities (dashed black lines), quantifying the energy-change statistics of the V-type three-level system subjected to incoherent light source, as a function of the dimensionless time $t\gamma/(2\pi)$. The quasiprobabilities refer to the (energy) transitions between the levels $|a\rangle,|b\rangle,|c\rangle$ of the system. Here, the imaginary parts of all the quasiprobabilities are equal to zero. For all the panels, we use the parameter setting at points (i)-(iv) with $p=-1$, and we distinguish between the contributions $q_{r,s}^{\rm diag}$ and $q_{r,s}^{\rm coh}$ depending respectively on $\chi_S$ (solid red lines) and ${\rm diag}(\hat{\rho}_S(0))$ (dash-dotted blue lines), where ${\rm diag}(\hat{\rho}_S(0)),\chi_S$ linearly decompose the initial density operator $\hat{\rho}_S$.
}
\label{fig:KDQ_extractable_work}
\end{figure}

Let us now show an example (with some plots) of these quantum behaviours involving the generation of Fano coherences. For this purpose, we take the following parameters' setting:
\begin{itemize}
\item[(i)]
V-type three-level system: Spontaneous decay rates (from $|a\rangle,|b\rangle$ to $|c\rangle$) $\gamma_a = \gamma_b \equiv \gamma \approx 3.0091 \cdot 10^{6}$ [rad/s]; energies $E_3 = \hbar\omega_a$, $E_2 = \hbar\omega_b$, $E_1 = \hbar\omega_c$ with $E_1 \leq E_2 \leq E_3$; $\omega_a = D + {\rm \Delta}/2$, $\omega_b = D - {\rm \Delta}/2$ and $\omega_c=0$, with $D \approx 0.785 \cdot 10^{9}$ [rad/s] (optical transition) and $\rm\Delta$ a fraction (10\%) of the spontaneous decay rate's value, \emph{i.e.} ${\rm\Delta} = 10\% \gamma = 0.1 \gamma$. In the figures we are going show below, the units of measurement of the plotted quantities are re-scaled such that $\hbar=1$.
\item[(ii)]
Incoherent source (sunlight radiation, or even noisy laser with quite larger emission bandwidth): Average photons number $\bar{n} = 3$; alignment parameter (between the dipole moments of the transitions $|a\rangle\leftrightarrow|c\rangle,|b\rangle\leftrightarrow|c\rangle$) $p=\cos\Theta=-1,-0.75,-0.5,-0.25$.  
\item[(iii)]
The bare Hamiltonian $\hat{H}_S$ of the V-type three-level system is proportional to the spin-1 operator 
\begin{equation*}
    S_z = \frac{ \hbar }{\sqrt{2}}\begin{pmatrix}
          0 & 1 & 0 \\
          1 & 0 & 1 \\
          0 & 1 & 0 
    \end{pmatrix}
\end{equation*}
along the $z$-axis. This means that the energy projectors $\hat{\Pi}_k$ resulting from its spectral decomposition are given by the outer product of the computational basis $|c\rangle \equiv (0,0,1)^T$, $|b\rangle \equiv (0,1,0)^T$ and $|a\rangle \equiv (1,0,0)^T$.
\item[(iv)]
Initial quantum state of the three-level system: $\hat{\rho}_S(0)=|\psi_0\rangle\!\langle\psi_0|$ with 
\begin{equation}\label{eq:initial_state_psi0}
|\psi_0\rangle = \alpha_a|a\rangle + \alpha_b \, e^{i\phi_b} |b\rangle + \alpha_c|c\rangle
\end{equation}
with $\alpha_a=\sqrt{0.3}$, $\alpha_b=\sqrt{0.3}$, $\phi_b=\pi$, and $\alpha_c=\sqrt{0.4}$; note that $\alpha_a^2 + \alpha_b^2 + \alpha_c^2 = 1$ to ensure probability conservation. As previously anticipated, the initial density operator of the three-level system (thus, at the beginning of the thermodynamic transformation under scrutiny) contains quantum coherence along the eigenbasis of $\hat{H}_S$. 
\end{itemize}

Now, using this parameters setting, we show two distinct plots: one concerning the average energy-change $\langle{\rm\Delta}E\rangle$ as a function of the dimensionless time $t\gamma/(2\pi)$ (Fig.~\ref{fig:extractable_work}), and the other regarding the underlying KDQ distribution (Fig.~\ref{fig:KDQ_extractable_work}).
For both plots we numerically solve the linear differential equations \eqref{eq:nineteen} that describe the dynamics responsible for the generation of Fano coherence in Markovian regime. The values of the parameters inserted in Eqs.~\eqref{eq:nineteen} are those provided at points (i)-(v) above. Moreover, we consider the results given by splitting the KDQ as in Eq.~\eqref{eq:linear_decomposition_KDQ}, where $\hat{\rho}_S(0)=|\psi_0\rangle\!\langle\psi_0|$ is linearly decomposed in two matrices ${\rm diag}(\hat{\rho}_S(0))$ and $\chi_S$ containing the diagonal and off-diagonal elements of $\hat{\rho}_S(0)$ respectively. 
We denote the two contributions of the KDQ 
\begin{equation}\label{eq:numerics_KDQ}
q_{r,s} = {\rm Tr}\Big[ |s\rangle\!\langle s|\,\Phi\big[ 
|r\rangle\!\langle r|\psi_0\rangle\!\langle\psi_0| \big]\Big] 
= \langle r|\psi_0\rangle \left\langle s| \,\Phi\big[ |r\rangle\!\langle\psi_0| \big] |s\right\rangle,
\end{equation}
with $r,s = a,b,c$, as $q_{r,s}^{\rm diag}$ and $q_{r,s}^{\rm coh}$ respectively. In \eqref{eq:numerics_KDQ}, the quantum map $\Phi[\cdot]$ is derived from equations of motion \eqref{eq:sol_x_t}-\eqref{eq:sol_z_t}.

From Fig.~\ref{fig:extractable_work} we evince that the ranges of parameters at points (i)-(v) are such that $\langle{\rm\Delta}E\rangle = 0$, as long as the initial density operator $\hat{\rho}_S$ of the three-level system does not contain quantum coherence $\chi_S$ (with respect to the basis diagonalizing $\hat{H}_S$). We stress that, by construction, such a result cannot be provided by the TPM scheme. On the contrary, by including quantum coherences as given by Eq.~\eqref{eq:initial_state_psi0}, $\langle{\rm\Delta}E\rangle = \langle{\rm \Delta}E\rangle_{\rm coh} \leq 0$ that entails a non-negligible amount of extractable work. Indeed, $\langle{\rm\Delta}E\rangle = \langle{\rm \Delta}E\rangle_{\rm diag} + \langle{\rm \Delta}E\rangle_{\rm coh}$ but $\langle{\rm \Delta}E\rangle_{\rm diag}=0$ in our case study. Thus, in Fig.~\ref{fig:extractable_work} we prove that even a quantum process that is driven by an incoherent source can generate extractable work, provided that the quantum system on which the process is realized is initialized in a superposition state of the energy eigenstates. Moreover, both the magnitude of $|\langle{\rm\Delta}E\rangle|$ and the time interval in which $|\langle{\rm\Delta}E\rangle| \neq 0$ can be linearly enhanced by increasing the value (with sign) of the alignment parameter $p \in [-1,1]$. Such an effect is maximized for $p=-1$, whereby ${\rm max}-\langle{\rm\Delta} E\rangle \approx 17\% \omega_a$ and remains at that value indefinitely as long as the incoherent light source is active. This finding is related (and thus consistent) with the already-known fact that $|p|=1$ entails quasi-stationary Fano coherences, ideally for an arbitrarily large time $t$~\cite{Koyu:Tscherbul,Koyu:Dodin,Dodin:Brumer}. It is worth noting the sign of $p$ is not relevant for the solution $\hat{\rho}_S(t)$ of the quantum system dynamics, but it matters for the sign of $\langle{\rm\Delta}E\rangle$ and thus for the nature of the thermodynamics process we are investigating. In fact, using the ranges of parameters at point (i)-(iv), $p$ negative entails extractable work, while $p$ positive means absorbed energy-change.

In the $9$ panels of Fig.~\ref{fig:KDQ_extractable_work} we plot the full distribution of KDQ (dashed black lines) $q_{r,s}$ with $r,s=a,b,c$. Such a quasiprobability distribution underlies the energy-change statistics and thus the extractable work in Fig.~\ref{fig:extractable_work}. In doing this, we use again the parameters setting at points (i)-(iv) but with $p=-1$, whereby the imaginary parts of all the plotted KDQ are equal to zero. In the figure, we distinguish between $q_{r,s}^{\rm diag}$ and $q_{r,s}^{\rm coh}$ of $q_{r,s}$, which we recall are the contributions stemming respectively from the matrices containing the diagonal and off-diagonal elements of $\hat{\rho}_S(0)$. We can observe that $q_{ac},q_{ab},q_{aa}$ have a contribution of $q_{r,s}^{\rm coh} \neq 0$ (solid red lines in the figure), which is due to initializing the system in a state with quantum coherence (with respect to the eigenbasis of $\hat{H}_S$). Notably, the quasiprobability $q_{ab}$ is globally negative in a transient time interval. In this regard, it is worth recalling that the Fano interference can arise between the excited levels $|a\rangle,\,|b\rangle$ of the three-level system. Hence, the presence of negativity in the corresponding KDQ describing energy-change fluctuations is an hallmark of Fano coherence generation occurring in a non-classical regime.
\begin{figure}[t]
\centering
\subfigure[][]{\includegraphics[width=0.45\textwidth]{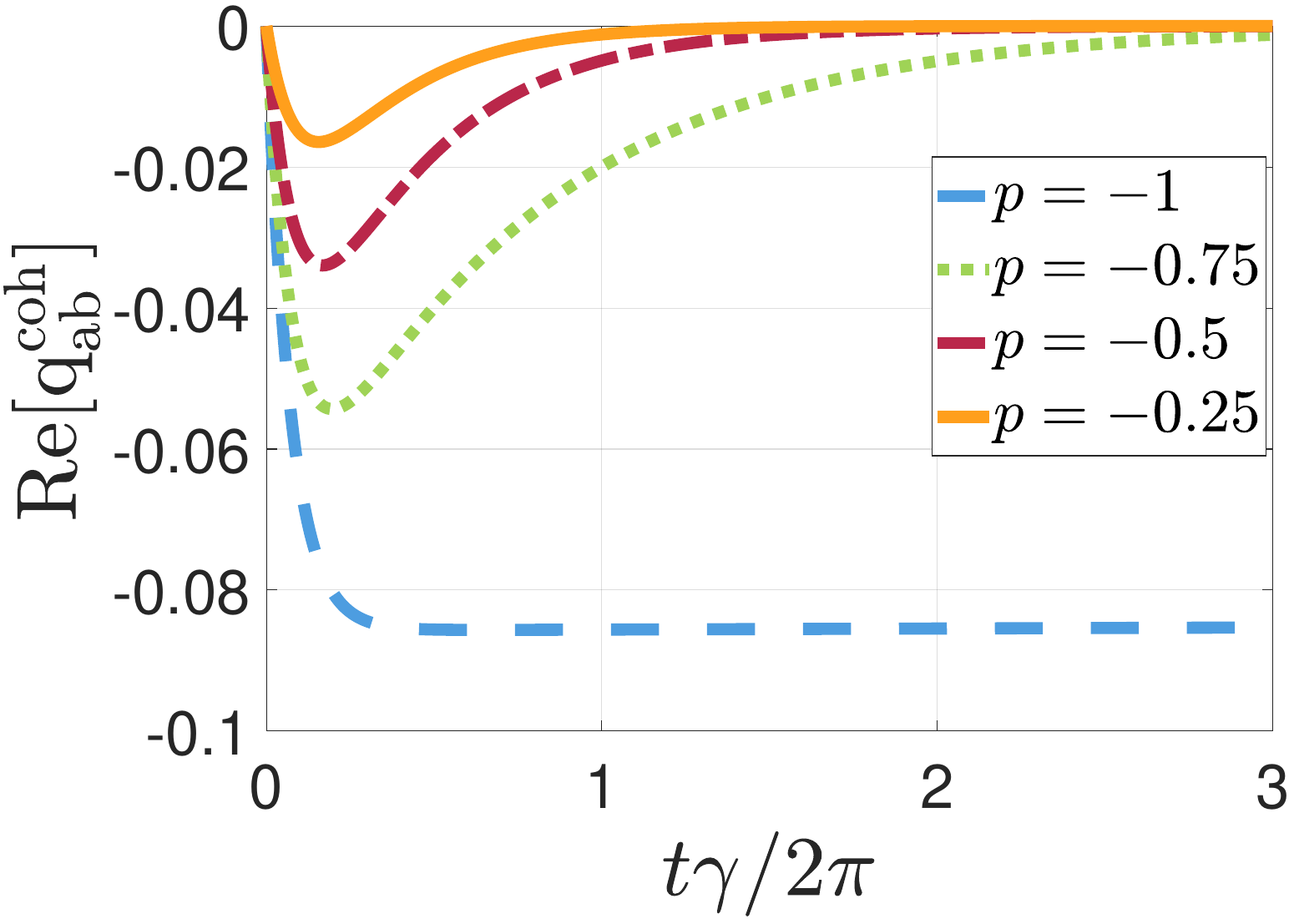}}
\hspace{1pt}
\subfigure[][]{\includegraphics[width=0.45\textwidth]{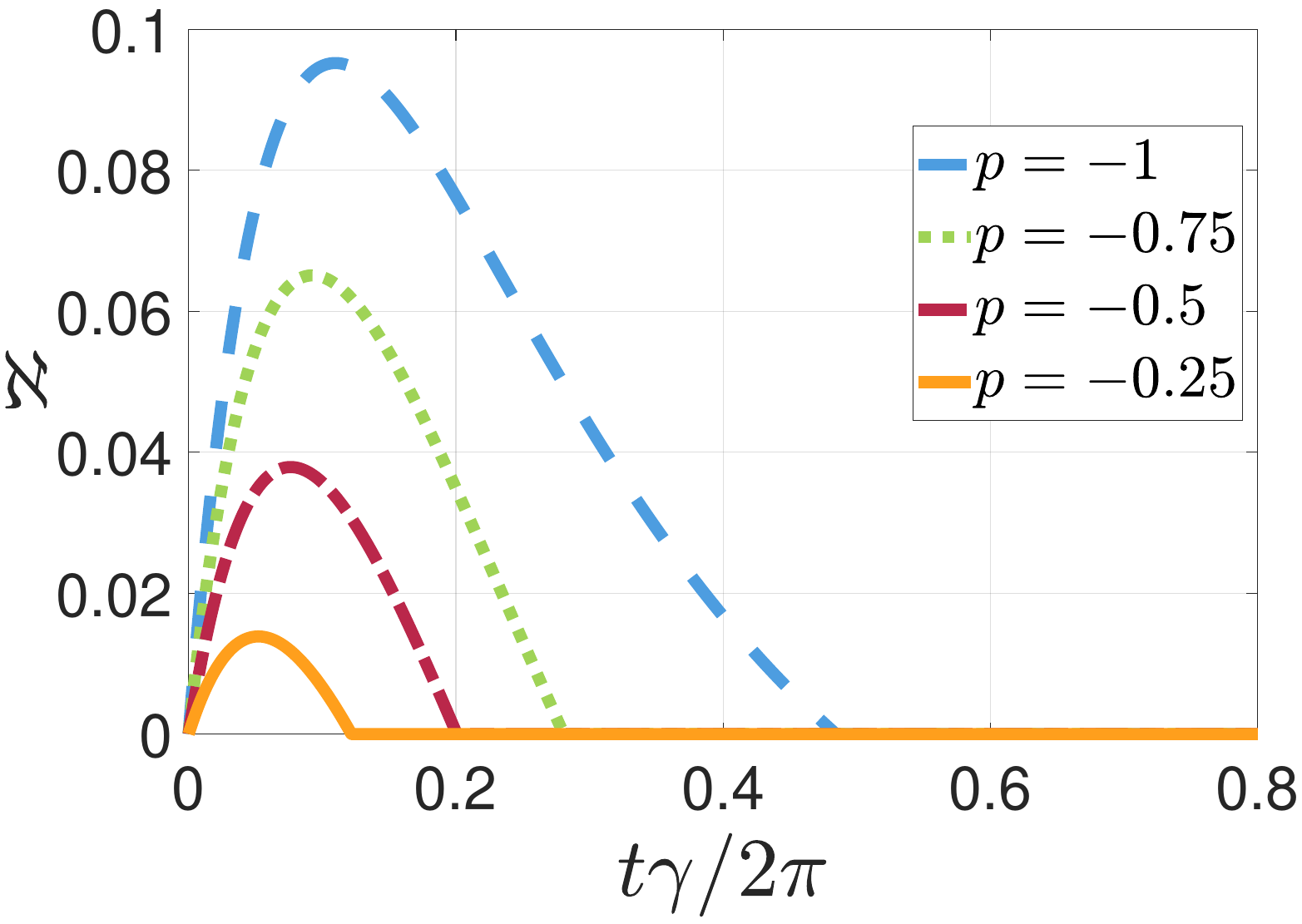}}
\caption{(a) Real part of $q_{a,b}^{\rm coh}$ [panel (a)] and 
the non-positivity functional $\aleph$ [panel (b)], as a function of the dimensionless time $t\gamma/(2\pi)$, for $p=-1,-0.75,-0.5,-0.25$. For both panels, the ranges of parameters at point (i)-(iv) are considered.}
\label{fig:negativity}
\end{figure}
We complete this analysis by showing in Fig.~\ref{fig:negativity} that: 
\begin{itemize}
\item[(i)] 
The real part of $q_{a,b}^{\rm coh}$ (plotted as a function of time) monotonically grows by increasing the value of the alignment parameter $p$ that effectively represents a control knob to enhance the negativity of the corresponding KDQ [panel (a)].
\item[(ii)]
The non-positivity functional $\aleph$ of the KDQ distribution of energy-changes is $>0$ in a transient time interval, at least in the parameters setting at points (i)-(iv). Interestingly, $\aleph$ is maximized for $p=-1$.
\end{itemize}

\subsection{Optimization of energy extraction}\label{sec_thermodynamics}

In the previous section, we have introduced a case study in which $\langle{\rm \Delta}E\rangle_{\rm diag}$, dependent on the diagonal elements of $\hat{\rho}_S(0)$, is zero for any time $t$. In this section our focus shifts to optimizing some key parameters of the model, including the initial quantum state of the three-level system, in order to maximize the amount of extractable energy. This optimization aims to maximize the value of $-\langle{\rm \Delta}E\rangle_{\rm coh}$ arising from the off-diagonal elements of $\hat{\rho}_S(0)$. As mentioned earlier, such an optimization also leads to an enhancement of negativity.

Achieving the condition $\langle{\rm \Delta}E\rangle_{\rm diag}=0$ relies solely on specific values of $\bar{n}$ and $\rho_{cc}(0) = |\alpha_c|^2$, under the assumption that the initial state of the system is given by Eq.~\eqref{eq:initial_state_psi0}. 
The analytical formula returning the values of $\bar{n},\rho_{cc}(0)$ such that $\langle{\rm \Delta}E\rangle_{\rm diag}=0$ is unknown. However, to attain $\langle{\rm \Delta}E\rangle_{\rm diag}=0$, we observe that increasing the value of $\bar{n}$ leads to the need to decrease $\rho_{cc}(0)$, and vice-versa. For instance, in the weak pumping regime ($\bar{n}<1$), the condition $\langle{\rm \Delta}E\rangle_{\rm diag}=0$ is satisfied for $\bar{n}=0.5$ and $\rho_{cc}(0)=0.6$. Conversely, in the strong pumping regime ($\bar{n}>1$), the condition $\langle{\rm \Delta}E\rangle_{\rm diag}=0$ holds for $\bar{n}=3$ and $\rho_{cc}(0)=0.4$ that are right the values already used in Sec.~\ref{sec_certification}. Choosing $\bar{n}$ above (below) the value allowing for $\langle{\rm \Delta}E\rangle_{\rm diag}=0$, for a given $\rho_{cc}(0)$, leads to $\langle{\rm \Delta}E\rangle_{\rm diag}$ being nonzero and either positive (negative). These considerations are valid for any values of $p$, but in what follows we specifically select $p=-0.5$. 
Once the condition $\langle{\rm \Delta}E\rangle=\langle{\rm \Delta}E\rangle_{\rm coh}$ is established, the optimization of $\langle{\rm \Delta}E\rangle_{\rm coh}$ is determined by the initial state $|\psi_0\rangle = \alpha_a \, e^{i\phi_a}|a\rangle + \alpha_b \, e^{i\phi_b} |b\rangle + \alpha_c\, e^{i\phi_c}|c\rangle$, where here we are considering a more general state featuring also the relative phases $\phi_a,\phi_c$ in addition to $\phi_b$.

\begin{figure}[!t]
\centering
\subfigure[]{\includegraphics[width=0.425\textwidth]{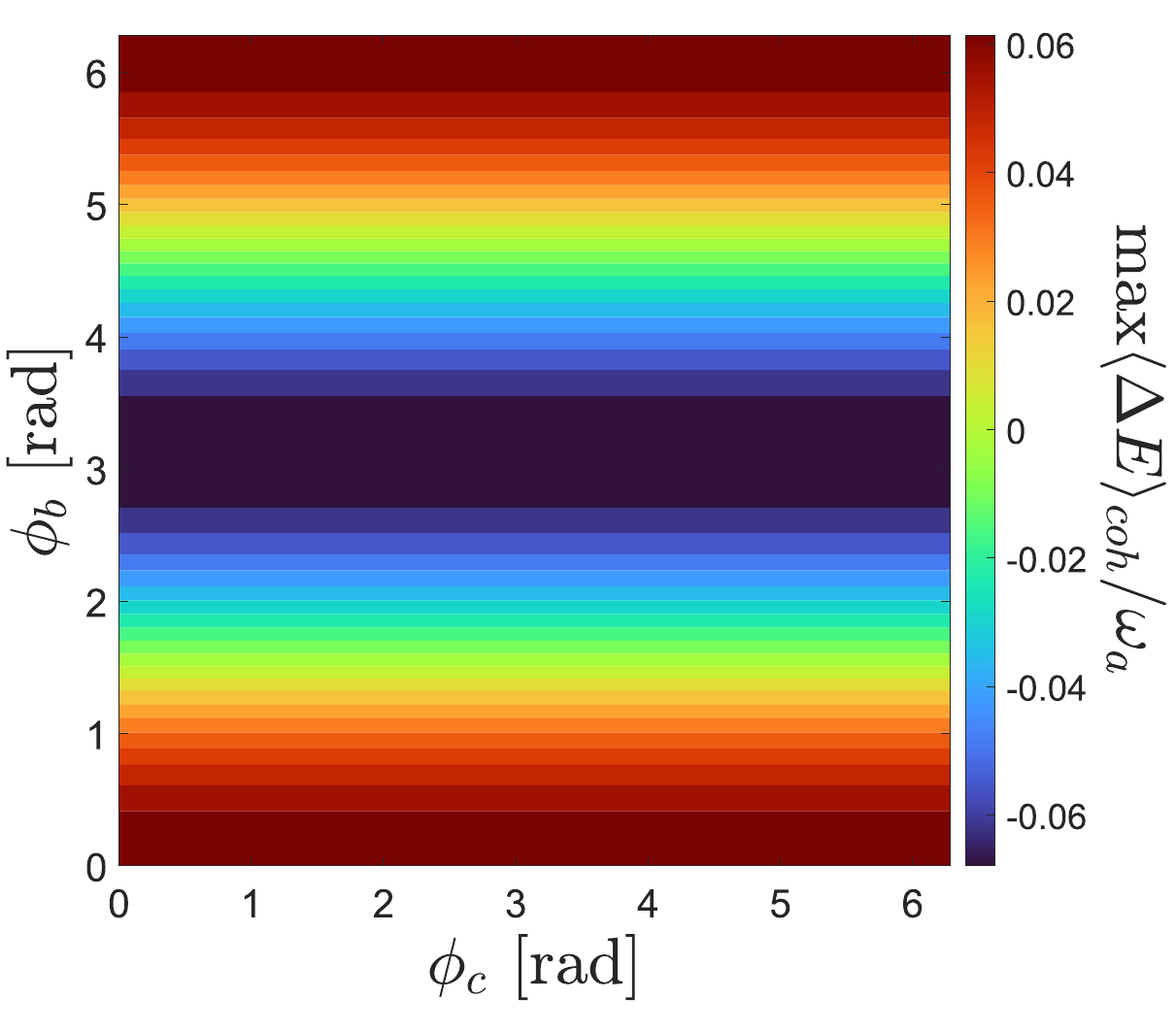}}
\hspace{1pt}
\subfigure[]{\includegraphics[width=0.425\textwidth]{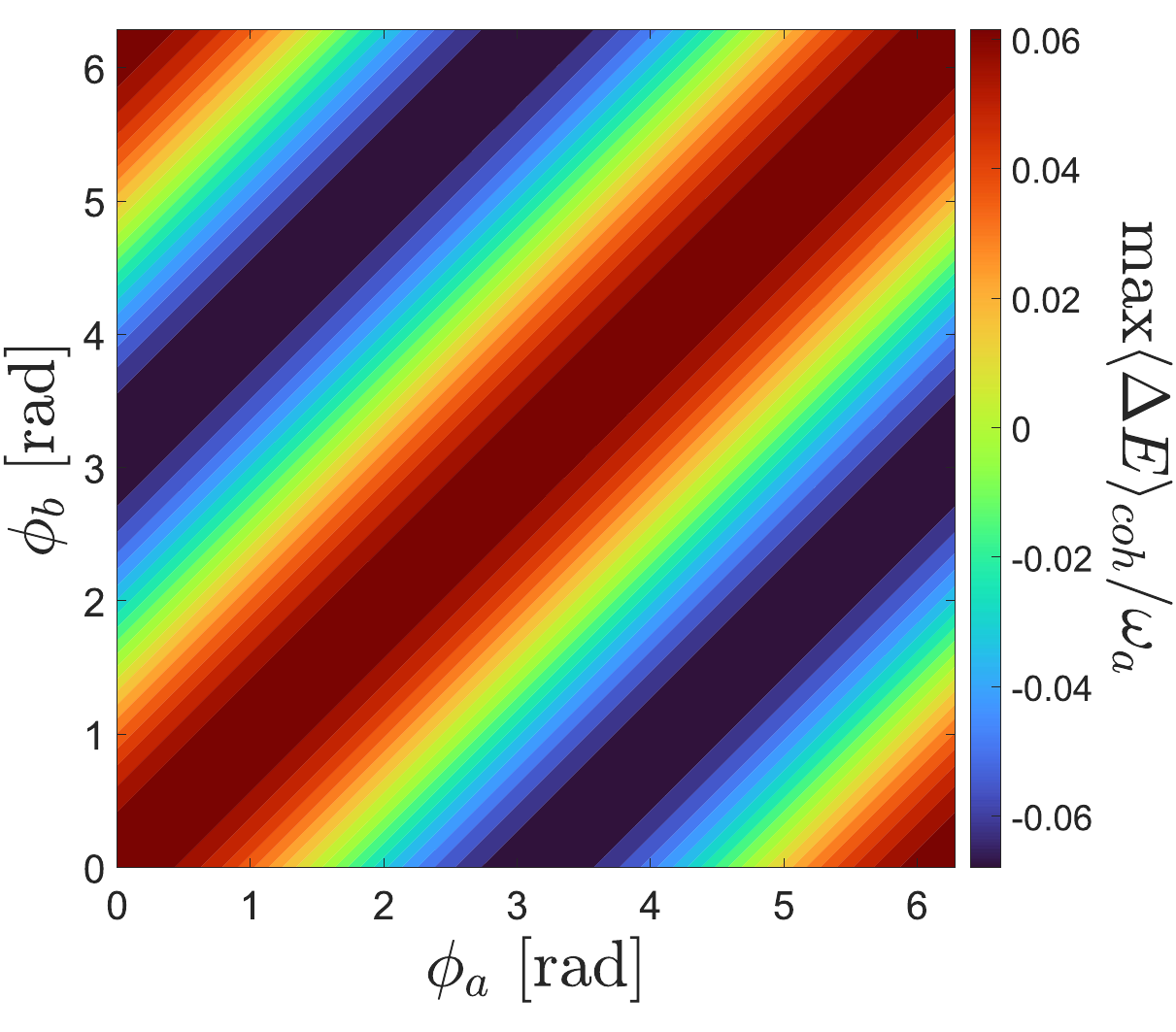}}
\caption{Largest values of $\langle {\rm \Delta} E \rangle_{\rm coh}$ including its sign, re-scaled by $\omega_a$, as a function of the relative phases $\phi_b,\phi_c$ [panel (a)], and $\phi_a,\phi_b$ [panel (b)]. In both panels the value of $p$ has been set to $-0.5$, $\bar n = 3$, $\rho_{cc}=0.4$, and $\rho_{aa}=\rho_{bb}=0.3$.}
\label{fig:colormap}
\end{figure}

Setting the values $\bar{n}=3$ and $\rho_{cc}(0)=0.4$, we take the populations $\rho_{aa}=|\alpha_a|^2=\rho_{bb}=|\alpha_b|^2=0.3$, and we vary the relative phases $\phi_a,\,\phi_b,\,\phi_c$ of $|\psi_0\rangle$ within the range $[0,2\pi]$. Interestingly, setting one of the relative phases to zero does not impact neither the maximum attainable value for $\langle{\rm \Delta}E\rangle$ nor the maximum extractable/absorbed work. Moreover, by using $\phi_a=0$ or $\phi_b=0$ and $\phi_c=0$, we identify two distinct scenarios that we are going analyze in detail:
\begin{itemize}
\item[(i)] ${\bm\phi_a=0}$ or ${\bm\phi_b=0}$:\\
Setting $\phi_a=0$, the two relative phase vary within the range $[0,2\pi]$, and we then record the corresponding values of $\langle{\rm \Delta}E\rangle$. Fig.~\ref{fig:colormap}a highlights the maximum values of $\langle{\rm \Delta}E\rangle_{\rm coh}$ by varying the value of the phases $\phi_b,\,\phi_c$. From the figure we observe that, in this setting, $\phi_c$ does not affect neither the magnitude nor the sign of $|\langle {\rm\Delta} E \rangle_{\rm coh}|$. Conversely, the relative phase $\phi_b$ significantly influences the quantity $|\langle {\rm\Delta} E \rangle_{\rm coh}|$. The magnitude $|\langle {\rm\Delta} E \rangle_{\rm coh}|$ is zero for $\phi_b=\pi/2$, and increases in both directions either towards $\phi_b = 0$ or $\phi_b = \pi$, but with opposite sign. The value $\phi_b=\pi$ represents a line of mirroring symmetry.
The results depicted in Fig.~\ref{fig:colormap}a are the same if we set $\phi_b=0$ instead of $\phi_a=0$ and we vary the relative phases $\phi_a,\,\phi_c$.
\item [(ii)] ${\bm\phi_c=0}$:\\
In this scenario we explore how the largest values of $\langle{\rm \Delta}E\rangle_{\rm coh}$, with sign, change by varying the values of the phases $\phi_a$ and $\phi_b$ across the range $[0,2\pi]$; see Fig.~\ref{fig:colormap}b.   
Unlike the symmetry observed in Fig.~\ref{fig:colormap}a, a different pattern emerges in Fig.~\ref{fig:colormap}b, whereby the mirroring symmetry line is given by the condition $\phi_a=\phi_b$.
\end{itemize}

We recall that in Fig.~\ref{fig:colormap} the value of $p$ has been set to $-0.5$. However, if one is free to also vary $p$, then we would observe that the sign of $p$ is responsible to change the sign of $\langle {\rm\Delta }E\rangle$, such that whenever $p<0$ the sign of $\langle {\rm\Delta }E\rangle$ is the same in Fig.~\ref{fig:colormap}, while for $p>0$ the condition is reversed. Similarly, the magnitude of $p$ is responsible to change the magnitude of $\langle{\rm\Delta }E\rangle$, such that decreasing the magnitude of $p$ decreases the largest value of $\langle{\rm\Delta} E\rangle$. We have previously noticed this behaviour also in Fig.~\ref{fig:extractable_work}.  
Before proceeding, it is also worth stressing that selecting $\phi_a=0,\,\phi_b=\pi,\,\phi_c=0$ in $|\psi_0\rangle$ leads to the maximization of the extractable work in Fig.~\ref{fig:colormap}.

\begin{figure}[t!]
\centering
\includegraphics[width=0.7\textwidth]{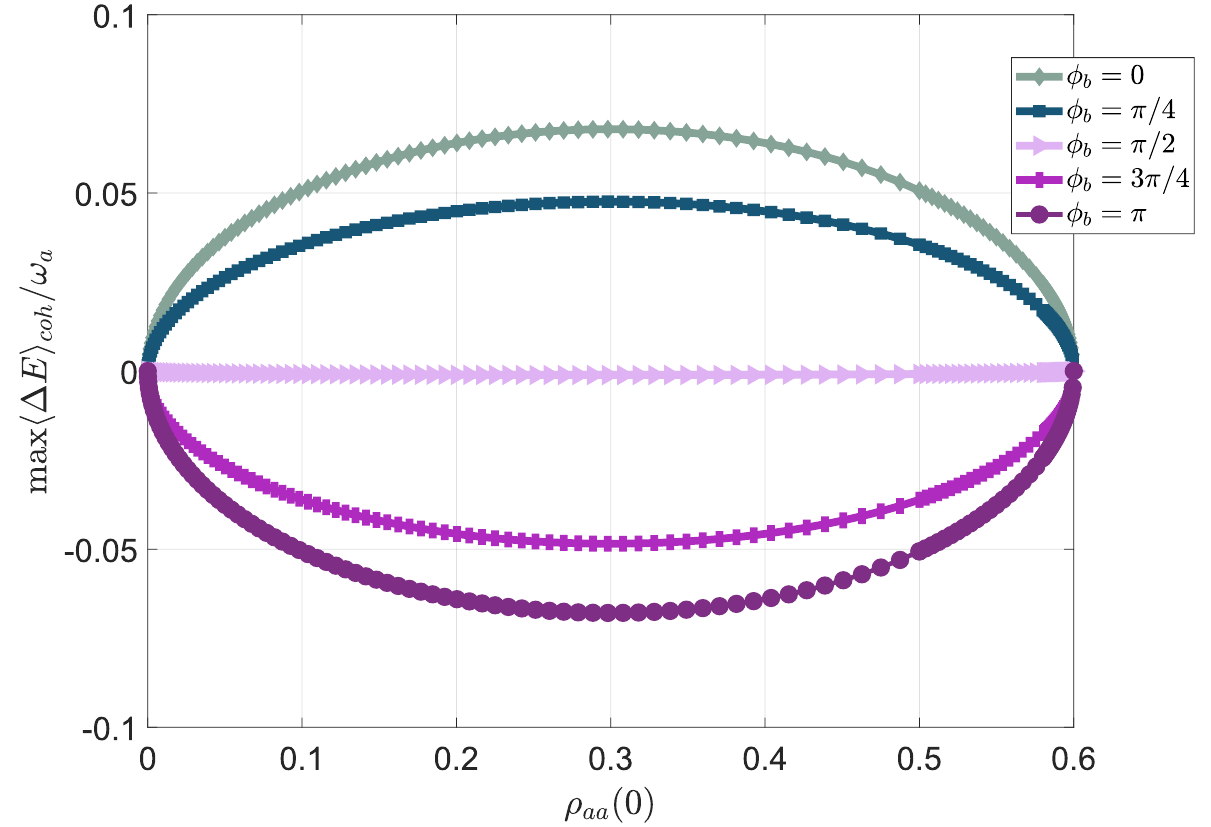}
\caption{Maximum value of $\langle {\rm\Delta} E \rangle_{\rm coh}$ as a function of $\rho_{aa}(0)$, initial population of the excited state $|a\rangle$. The ground state is fixed at $\rho_{cc}(0) = 0.4$ and the relative phases are $\phi_a=\phi_c=0$, while $\rho_{bb}(0) = 1-\rho_{cc}(0)-\rho_{aa}(0)$. The alignment parameter $p$ is set to $p=-0.5$. Different values of $\phi_b$ are taken into account.}
\label{fig:pop_variation}
\end{figure}

Let us now analyze how $\langle{\rm\Delta}E\rangle$ changes for different values of the populations $\rho_{aa}(0)$ and $\rho_{bb}(0)$, which pertain to the excited states $|a\rangle$ and $|b\rangle$. We do not directly consider $\rho_{cc}(0)$ (population in the ground level $|c\rangle$), as it is predetermined by $\bar{n}$. For instance, in the scenario with $p=-0.5,\bar{n} = 3 \Rightarrow \rho_{cc}(0)=0.4$, we vary only the value of the population $\rho_{aa}(0)$; indeed, $\rho_{bb}(0)$ changes according to the constraint $\rho_{bb}(t) = 1 - \rho_{aa}(t) - \rho_{cc}(t),\,\forall\,t$. The results depicted in Fig.~\ref{fig:pop_variation} illustrate ${\rm max}\langle{\rm\Delta}E\rangle_{\rm coh}$ as a function of $\rho_{aa}(0)$, with $\phi_b=0,\pi/4,\pi/2,3\pi/4,\pi$.
While $\rho_{cc}(0)$ may affect $\langle {\rm\Delta }E \rangle_{\rm diag}$, the initial populations $\rho_{aa}(0),\rho_{bb}(0)$ of the excited states impact $\langle {\rm\Delta} E \rangle_{\rm coh}$. Specifically, $\langle {\rm\Delta} E \rangle_{\rm coh}$ is zero when the three-level system is initialized with one among $\rho_{aa}(0),\rho_{bb}(0)$ is set to zero. Additionally, we observe that the maximum value of $\langle{\rm\Delta} E \rangle_{\rm coh}$ is obtained when $\rho_{aa}(0)=\rho_{bb}(0)$. The imbalance in favor of one over the other decreases ${\rm max}\langle{\rm\Delta}E\rangle_{\rm coh}$. As in Fig.~\ref{fig:colormap}a, varying $\phi_b$ from $0$ to $\pi$ enables a transition from the condition of maximum absorbed energy-change ($\phi_b=0$) to maximum extractable work ($\phi_b=\pi$), passing through a regime where $\langle {\rm\Delta}E\rangle_{\rm coh}=\langle {\rm\Delta} E \rangle_{\rm diag}=\langle {\rm\Delta}E\rangle=0$ ($\phi_b=\pi/2$). 

To sum-up, the optimal initial state configuration is achieved by setting the populations $\rho_{aa}(0),\rho_{bb}(0) \neq 0$ and $\rho_{aa}(0)=\rho_{bb}(0)$, while the value of $\rho_{cc}(0)$ is dictated by the $\bar{n}$ that allows for $\langle{\rm\Delta}E\rangle=\langle {\rm\Delta} E \rangle_{\rm coh}$. Finally, regarding the relative phases $\phi_a,\phi_b,\phi_c$ entering the initial wave-function $|\psi_0\rangle$, setting all the three to zero means maximum absorbed energy-change, whereas choosing $\phi_b=\pi$ (with $\phi_a=\phi_c=0$) entails maximum extractable work.

\subsection{Efficiency of the process}

The assessment of the thermodynamic efficiency is crucial in any energy conversion process, to gauge the performance in transforming a form of energy (the input energy $\mathcal{E}_{\rm in}$) in another (extractable energy $\mathcal{E}_{\rm extr}$) for practical uses. The efficiency is generally defined as follows:
\begin{equation}
\eta \equiv \frac{\mathcal{E}_{\rm extr}}{\mathcal{E}_{\rm in}}\,.
\end{equation}
In our case study, as introduced in Sec.~\ref{sec_certification}, the extractable work is given by the quantity $-\langle{\rm \Delta}E(t)\rangle>0$, where only the contribution from the off-diagonal elements of $\hat{\rho}_S$ accounts. Conversely, the energy that drives the system, which originates from the incoherent field, is  $\mathcal{E}_{\rm in} = \bar{n}\hbar\omega_{ac}$ that is the average energy of the photons impinging on the system. Hence,
\begin{equation}\label{eq:efficiency}
\eta(t) = \frac{-\langle{\rm \Delta}E(t)\rangle}{\bar{n}\hbar\omega_{ac}} \,.
\end{equation}
Eq.~\eqref{eq:efficiency} reveals that the time dependence of the efficiency follows the one of $\langle{\rm \Delta}E\rangle$ depicted in Fig.~\ref{fig:extractable_work}. Consequently, the efficiency reaches its peak when $\langle{\rm \Delta}E\rangle$ is maximized, which occurs at a specific instant $t$ that we denote as $\widetilde{t}$. Notably, in the scenario with $p=-1$, both $\eta$ and $\langle{\rm \Delta}E\rangle$ attain a maximum stationary value.

Based on the optimization analysis in Sec.~\ref{sec_thermodynamics}, we focus on the condition yielding the maximum extractable energy, given by $\rho_{aa}(0)=\rho_{bb}(0)=0.3$ with $\rho_{cc}(0)=0.4$, $\bar{n}=3$ and $\phi_{a}=\phi_{c}=0,\phi_b=\pi$. In Tab.~\ref{tab1} we present the achievable maximum efficiency together with the time instants at which it is obtained, for various values of $p$.
\begin{table}[t!]
    \centering
    \begin{tabular}{ccccc}
        \toprule
        p & \qquad & $\eta_{\rm max}$ & \qquad & $\displaystyle{\frac{\widetilde{t}\gamma}{2\pi}}$ \\
        \midrule
        -1 & \qquad & 6\% & \qquad & $>0.40$ \\
        -0.75 & \qquad & 4\% & \qquad & 0.20 \\
        -0.5 & \qquad & 2\% & \qquad & 0.17 \\
        -0.25 & \qquad & 1\% & \qquad & 0.16 \\
        \bottomrule
        \vspace{0.05cm}
    \end{tabular}
    \caption{}
    \label{tab1}
\end{table}        
We conclude by noting that we have not inserted, among the costs in the calculation of the efficiency, the energy for preparing the initial state of the three-level system. This is because we are implicitly assuming to work in a condition where the preparation of a superposition of Hamiltonian eigenstates as the initial state is given for granted. However, this assumptions could be rectified when dealing with the experimental realization of a process for Fano coherence generation.

\section{Discussion}

In this paper we discuss the energetics behind the generation of Fano-like quantum coherence, by using a prototypical V-type three-level system (finite - dimensional quantum system) in interaction with an incoherent radiation field. The latter is assumed as consisting of a continuum of modes, shaped on a broadband frequency range. If the excited levels of the three-level system are taken close enough, then Fano coherences develop for a transient time interval. They become stationary in the limiting case the excited states are degenerate. Thus, the following question arises: ``To what extent the process generating Fano coherence can be considered genuinely quantum?'' The answer to this question would constitute a first attempt to certify the quantumness of a process, driven by an incoherence field, while inducing quantum effects in a nonequilibrium regime.

For this purpose, we here determine the Kirkwood-Dirac quasiprobability distribution of the (time-dependent) energy-changes in the three-level system under scrutiny, while the incoherent radiation field is active. If the real part of some quasiprobability is negative, or even some quasiprobability is complex, then one can witness the onset of a genuine quantum effect linked to quantum interference profiles. Necessary condition for that is the non-commutativity of the initial state of the quantum system with the Hamiltonian $\hat{H}_S$ at the beginning of the dynamical transformation (recall that in the process generating Fano coherence, the Hamiltonian is time-independent). Thus, as expected, we observed that initializing the three-level system in a superposition of the Hamiltonian eigenstates, there exist a range of parameters in which negative quasiprobabilities arise but still with zero imaginary parts. 
Initializing in the ground state of $\hat{H}_S$ does not lead to the same result.

Under the same parameter setting and the same choice of the initial state ($\bar{n}=3$ and $\rho_{cc}(0)=0.4$), we find that $\langle{\rm\Delta}E\rangle = \langle{\rm \Delta}E\rangle_{\rm coh} \leq 0$ in a given time interval for any value of the alignment parameter $p$, except $p=0$. Interestingly, albeit the input light source is incoherent, the maximum efficiency of the thermodynamic process goes up to $6\%$, and becomes stationary for $p=-1$. This findings motivates us in further investigating the design and optimization of a (coherent or incoherent) coupling with an external load that can act as energy battery~\cite{CampaioliReview2023} or quantum flywheel~\cite{LevyPRA2016}.

The results we provide in this paper could be experimentally validated via inferring the real part of the quasiprobabilities for the energy-change statistics. For this purpose, as recently shown in Refs.~\cite{hernandez2022experimental,LostaglioKirkwood2022}, we can resort to reconstruction procedures, either entirely based on projective measurements or implementing an interferometric scheme.
Even the experimental realization of a V-type three-level system conducive to Fano coherence is achievable. This can be implemented using an atomic platform comprising a gas of a suitable atomic species maintained at a constant temperature. Choosing a cold or hot gas can be relevant for
the preparation of the initial state of the quantum system in a superposition of the Hamiltonian eigenstates. Indeed, lower is the temperature, and better should be the tunability of the parameters inducing state changes. Finally, the generation of Fano coherence may necessitate the polarization of the incoherent radiation field, a requirement that varies depending on the selected atomic species~\cite{Dodin:Tscherbul,Koyu:Dodin}.

\section{Methods}
\subsection{Explicit computation of the double commutator in the differential equation of $\hat{\rho}_S$}\label{appendix_model}

In this appendix we report the complete derivation of the double commutator in the differential equation of $\hat{\rho}_S$. For this purpose, let us take the second term on the right-hand
side of Eq.~(\ref{eq:four}). By using Eqs.~\eqref{eq:interaction_H}-(\ref{eq:def1}) and further expanding the double commutator in Eq.~(\ref{eq:four}), we derive terms of the form
\begin{eqnarray}\label{eq:double_commutator_1}
&&\int_0^t \sum_{\lambda,\lambda'} \sum_{\bf{k},\bf{k'}} \left(g_{\bf{k},\lambda}^{r}\right)^*g_{\bf{k'},\lambda'}^{r}\,e^{-i(\omega_{rc}-\nu_k)t+i(\omega_{rc}-\nu_{k'})t'}\Big[\langle\hat{a}_{\bf{k},\lambda}^\dagger\hat{a}_{\bf{k'},\lambda'}\rangle\left( \hat{\sigma}_{rc}^{-}\hat{\sigma}_{rc}^{+}\hat{\rho}_S(t')-\sigma_{rc}^+\rho_S(t')\sigma_{rc}^-\right)\nonumber \\
&& + \langle\hat{a}_{\bf{k'},\lambda'}\hat{a}_{\bf{k}\lambda}^\dagger\rangle\left(\hat{\rho}_S(t')\hat{\sigma}_{rc}^{+}\hat{\sigma}_{rc}^{-} - \hat{\sigma}_{rc}^{-}\hat{\rho}_S(t')\hat{\sigma}_{rc}^+\right)\Big] dt' 
\end{eqnarray}
with $r=a,b$. Moreover, also the following crossing terms (namely involving both the levels $|a\rangle$ and $|b\rangle$) arise: 
\begin{eqnarray}\label{eq:double_commutator_2}
&&\int_0^t \sum_{\lambda,\lambda'} \sum_{\bf{k},\bf{k'}} g_{\bf{k},\lambda}^{a}\left(g_{\bf{k'},\lambda'}^{b}\right)^{*}e^{i(\omega_{ac}-\nu_k)t-i(\omega_{bc}-\nu_{k'})t'}\Big[\langle\hat{a}_{\bf{k'},\lambda'}^\dagger\hat{a}_{\bf{k},\lambda}\rangle\left(\hat{\rho}_S(t')\hat{\sigma}_{bc}^-\hat{\sigma}_{ac}^{+}- \hat{\sigma}_{ac}^{+}\hat{\rho}_S(t')\hat{\sigma}_{bc}^-\right)\nonumber \\
&& +
\langle\hat{a}_{\bf{k},\lambda}\hat{a}_{\bf{k'},\lambda'}^\dagger\rangle\left(\hat{\sigma}_{ac}^+\hat{\sigma}_{bc}^-\hat{\rho}_S(t') - \hat{\sigma}_{bc}^-\hat{\rho}_S(t')\hat{\sigma}_{ac}^+\right)\Big] dt'\,.
\end{eqnarray}
Hence, from substituting the expectation values in Eqs.~(\ref{eq:five})-(\ref{eq:seven}), Eqs.~\eqref{eq:double_commutator_1}-\eqref{eq:double_commutator_2} simplify as
\begin{eqnarray}\label{eq:eight}
&&\int_0^t \sum_{\lambda} \sum_{\bf{k}} |g_{\bf{k},\lambda}^{r}|^2e^{-i(\omega_{rc}-\nu_k)(t-t')}\Big[\bar{n}_k\left(\hat{\sigma}_{rc}^-\hat{\sigma}_{rc}^+\hat{\rho}_S(t')-\hat{\sigma}_{rc}^+\hat{\rho}_S(t')\hat{\sigma}_{rc}^-\right)+\nonumber \\
&& + (\bar{n}_k+1)\left( \hat{\rho}_S(t')\hat{\sigma}_{rc}^+\hat{\sigma}_{rc}^{-} - \hat{\sigma}_{rc}^{-}\hat{\rho}_S(t')\hat{\sigma}_{rc}^+ \right)\Big] dt' 
\end{eqnarray}
and
\begin{eqnarray}\label{eq:nine}
&&\int_0^t \sum_{\lambda} \sum_{\bf{k}} g_{\bf{k},\lambda}^{a}\left(g_{\bf{k},\lambda}^{b}\right)^{*}e^{i(\omega_{ac}-\nu_k)t-i(\omega_{bc}-\nu_{k})t'}\Big[\bar{n}_k\left(\hat{\rho}_S(t')\hat{\sigma}_{bc}^-\hat{\sigma}_{ac}^+ - \hat{\sigma}_{ac}^+\hat{\rho}_S(t')\hat{\sigma}_{bc}^-\right)+\nonumber \\
&& + (\bar{n}_k+1)\left(\hat{\sigma}_{ac}^+\hat{\sigma}_{bc}^-\hat{\rho}_S(t') - \hat{\sigma}_{bc}^-\hat{\rho}_S(t')\hat{\sigma}_{ac}^+\right)\Big] dt'\,.
\end{eqnarray}

At this point we apply the \textit{Weisskopf-Wigner approximation} that assumes the all the frequency modes of the radiation field are closely spaced within a spherical volume. Also the fact that the radiation field is contained in a sphere is an approximation that helps to simplify the mathematical treatment of the model. However, it just leads to a small approximation error since the modes of the radiation fields are uncorrelated to each other, given that the (light) source is incoherent. The Weisskopf-Wigner approximation is formally provided by the replacement
\begin{equation}\label{eq:ten}
\sum_{\bf{k}}\longrightarrow \frac{2V}{(2\pi)^3}\int_0^{2\pi}d\phi\int_{0}^\pi d\theta\sin\theta\int_0^\infty dk \, k^2 \quad \text{with} \quad k \equiv \frac{ \nu_k }{c}
\end{equation}
($c$ is the speed of light), whose function indeed is to shift the discrete distribution of the radiation modes to a continuous distribution that we represent in spherical coordinates. Thus, implementing the Weisskopf-Wigner approximation \eqref{eq:ten} to Eqs.~\eqref{eq:eight}-\eqref{eq:nine} and using the definition of the coupling terms $g_{\bf{k},\lambda}^r$ of Eq.~\eqref{eq:def1} leads us to
\begin{equation}\label{eq:eleven}
\begin{split}
& \frac{1}{\hbar8\pi^3\varepsilon_0c^3}\int_0^t \int_0^{2\pi}d\phi \int_0^{\pi}\left|\bm{\mu}_{rc}\right|^2d\theta \sin\theta\int_0^\infty e^{-i(\omega_{rc}-\nu_k)(t-t')}\Big[ \bar{n}_k \left( \hat{\sigma}_{rc}^-\hat{\sigma}_{rc}^+\hat{\rho}_S(t') 
- \hat{\sigma}_{rc}^+\hat{\rho}_S(t')\hat{\sigma}_{rc}^-\right) + \\
&+(\bar{n}_k+1)\left(\hat{\rho}_S(t')\hat{\sigma}_{ac}^+\hat{\sigma}_{ac}^- - \hat{\sigma}_{ac}^-\hat{\rho}_S(t')\hat{\sigma}_{ac}^+\right)\Big]\nu_k^3 d\nu_k dt' 
\end{split}
\end{equation}
and
\begin{equation}\label{eq:twelve}
\begin{split}
& \frac{1}{\hbar8\pi^3\varepsilon_0c^3}\int_0^t \int_0^{2\pi}d\phi \int_0^{\pi}\left(\bm{\mu}_{ac}\cdot\bm{\mu}_{bc}\right)d\theta\sin\theta\int_0^\infty e^{i(\omega_{ac}-\nu_k)t-i(\omega_{bc}-\nu_{k})t'}\Big[\bar{n}_k \left(\hat{\rho}_S(t')\hat{\sigma}_{bc}^-\hat{\sigma}_{ac}^{+} - \hat{\sigma}_{ac}^+\hat{\rho}_S(t')\hat{\sigma}_{bc}^-\right) +\\
& + (\bar{n}_k+1)\left(\hat{\sigma}_{ac}^+\hat{\sigma}_{bc}^-\hat{\rho}_S(t') - \hat{\sigma}_{bc}^-\hat{\rho}_S(t')\hat{\sigma}_{ac}^+\right)\Big]\nu_k^3 d\nu_k dt' \,,
\end{split}
\end{equation}
where $\nu_k = k c$, and we made use of the assumption that the radiation is \textit{isotropic} (i.e., the modes of the radiation are uniformly distributed along all the spatial directions) and \textit{unpolarised}, \emph{i.e.},
\begin{equation}
\sum_\lambda\sum_{\bf{k}}g_{\bf{k},\lambda}^{r}g_{\bf{k},\lambda}^{s}=\bm{\mu}_{rc}\cdot\bm{\mu}_{sc}\frac{\nu_k}{2\varepsilon_0V}
\end{equation}
with $r,s=a,b$. When $\nu_k\neq\omega_{rc}$, the exponential terms in Eqs.~\eqref{eq:eleven}-\eqref{eq:twelve} oscillate rapidly such that they can be neglected. This means that we can effectively assume $\nu_k$ as an approximately constant quantity $\forall\, k$. In particular, we can consider that $\nu_k \approx \omega_{ac} \simeq \omega_{bc}$ for any $k$, given that in optical transitions $\omega_{ac}-\omega_{bc}=\rm{\Delta} \ll \omega_{ac},\omega_{bc}$. The result of this approximation in Eqs.~\eqref{eq:eleven}-\eqref{eq:twelve} is to set the following integral computations:
\begin{eqnarray*}
    &&\int_0^\infty e^{-i(\omega_{rc}-\nu_k)(t-t')}\nu_k^3 d\nu_k \approx e^{-i\omega_{rc}(t-t')}\int_{-\infty}^{\infty}e^{iv_k(t-t')}\nu_k^3 d\nu_k \approx \\
    &&\approx e^{-i\omega_{rc}(t-t')}\omega_{rc}^{3}\int_{-\infty}^{\infty}e^{iv_k(t-t')}d\nu_k = 2\pi\omega_{rc}^{3} e^{-i\omega_{rc}(t-t')}\delta(t-t')
\end{eqnarray*}
and 
\begin{eqnarray*}
    &&\int_0^\infty e^{i(\omega_{ac}-\nu_k)t -i(\omega_{bc}-\nu_k)t')}\nu_k^3 d\nu_k \approx e^{i\omega_{ac}(t-t')}\int_{-\infty}^{\infty}e^{-iv_k(t-t')}\nu_k^3 d\nu_k \approx \\
    &&\approx e^{i\omega_{ac}(t-t')}\omega_{ac}^{3}\int_{-\infty}^{\infty}e^{-iv_k(t-t')}d\nu_k = 2\pi\omega_{ac}^{3} e^{i\omega_{ac}(t-t')}\delta(t-t'),
\end{eqnarray*}
where in both equations we exploited the Fourier representation of the Dirac delta function. In this way, Eqs.~\eqref{eq:eleven}-\eqref{eq:twelve} simplify as
\begin{equation}\label{eq:thirteen}
\begin{split}
& \frac{\omega_{rc}^3\left|\bm{\mu}_{ac}\right|^2}{\hbar6\pi^2\varepsilon_0c^3} \int_0^t \Big[\bar{n}\left(\hat{\sigma}_{ac}^-\hat{\sigma}_{ac}^+\hat{\rho}_S(t')- \hat{\sigma}_{ac}^+\hat{\rho}_S(t')\hat{\sigma}_{ac}^-\right) + \\
&+(\bar{n}+1)\left(\hat{\rho}_S(t')\hat{\sigma}_{ac}^+\hat{\sigma}_{ac}^- - \hat{\sigma}_{ac}^-\hat{\rho}_S(t')\hat{\sigma}_{ac}^+\right)\Big]e^{-i\omega_{rc}(t-t')}2\pi\delta(t-t')dt'
\end{split}
\end{equation}
and
\begin{equation}\label{eq:fourteen}
\begin{split}
& \frac{\omega_{ac}^3\left(\bm{\mu}_{ac}\cdot\bm{\mu}_{bc}\right)}{\hbar6\pi^2\varepsilon_0c^3}\int_0^t 
\Big[ \bar{n}\left(\hat{\rho}_S(t')\hat{\sigma}_{bc}^-\hat{\sigma}_{ac}^+ 
 - \hat{\sigma}_{ac}^+\hat{\rho}_S(t')\hat{\sigma}_{bc}^-\right) +\\
& + (\bar{n}+1)\left(\hat{\sigma}_{ac}^+\hat{\sigma}_{bc}^-\hat{\rho}_S(t') - \hat{\sigma}_{bc}^-\hat{\rho}_S(t')\hat{\sigma}_{ac}^+\right) \Big] e^{i\omega_{ac}(t-t')}2\pi\delta(t-t')dt'
\end{split}
\end{equation}
with $\bar{n} = [\exp(\hbar\omega_{ac}/k_BT)-1]^{-1}$.

\subsection*{Acknowledgments}
We wish to acknowledge financial support from the project PRIN 2022 Quantum Reservoir Computing (QuReCo), and the PNRR MUR project PE0000023-NQSTI financed by the European Union--Next Generation EU.

\subsection*{Authors Contribution}
S.G. conceived and supervised the whole project; L.D. performed the theoretical calculation and numerical simulations with input from F.S.C. and S.G.; L.D. and S.G. carried out the analysis of the results. All authors contributed to the discussion, and the writing of the manuscript.




\end{document}